\documentclass[a4paper,12pt]{article}
\usepackage{chngcntr} 
\counterwithin{equation}{section} 

\usepackage[dvipdfmx]{graphicx}
\usepackage{authblk}
\usepackage{bm}
\usepackage{color}
\usepackage{array}
\usepackage{fancyhdr}
\usepackage{enumitem}
\usepackage{hyperref}
\usepackage{amsmath,amssymb}
  \usepackage{tabularx}
  \usepackage{cite}
  \usepackage{hhline}
   \newcolumntype{C}{>{\centering\arraybackslash}X}
   \newcolumntype{L}{>{\raggedright\arraybackslash}X}
   \newcolumntype{R}{>{\raggedleft\arraybackslash}X}
\setlistdepth{10}

\usepackage{titlesec}

\titleformat{\section}  
  {\fontsize{14}{16}\bfseries} 
  {\Roman{section}.} 
  {0.5em} 
  {} 
  [] 

\titleformat{\subsection}  
  {\fontsize{12}{14}\bfseries} 
  {\Alph{subsection}.} 
  {0.5em} 
  {} 
  [] 

  \titleformat{\subsubsection}  
  {\fontsize{10}{12}\bfseries} 
  {\arabic{subsubsection}.} 
  {0.5em} 
  {} 
  [] 

\usepackage[normalem]{ulem}

\newcommand{\ii}{\mathrm{i}}
\newcommand{\dd}{\mathrm{d}}

\newcommand{\rmc}{{\rm c}}
\newcommand{\del}{\partial}
\newcommand{\ee}{{\rm e}}

\definecolor{DarkBlue}{rgb}{0,0,0.7} 

\definecolor{DarkRed}{rgb}{0.65,0,0}

\newcommand{\repnuma}{NU-QG-12}
\newcommand{\repnumb}{RUP-25-22}
\newcommand{\emaila}{\sf yoo.chulmoon.k6@f.mail.nagoya-u.ac.jp}

\fancypagestyle{titlepage}{%
  \fancyhf{}
  \fancyhead[R]{%
    \repnuma\\
    \repnumb
  }
   \fancyfoot[L]{%
    \rule{0.2\textwidth}{0.4pt} \\[-\baselineskip]%
    \vspace{4mm}
    \emaila}

}

\title{\bf \large Ringdown in Vaidya spacetimes: time-dependent frequencies, Penrose limit and time-domain analyses}

\author[1,2]{\normalsize Chul-Moon~Yoo}
\author[3,4]{\normalsize Masashi~Kimura}
\author[1,2]{\normalsize Akihiro Ishibashi}
\author[1]{\normalsize Rikuto Ohashi}

\affil[1]{\small \em Graduate School of Science, Nagoya University, Nagoya 464-8602, Japan}
\affil[2]{\small \em Kobayashi-Maskawa Institute for the Origin of Particles and the Universe (KMI), \protect\\ Nagoya 464-8602, Japan}
\affil[3]{\small \em Department of Information, Artificial Intelligence and Data Science, \protect\\ Daiichi Institute of Technology, Tokyo 110-0005, Japan}
\affil[4]{\small \em Department of Physics, Rikkyo University, Toshima, Tokyo 171-8501, Japan}

\begin{document}
\baselineskip5.5mm
\date{}
\maketitle
\thispagestyle{titlepage}













\begin{abstract}
\baselineskip5.5mm
\normalsize
We examine the possible characterization of ringdown waves in a dynamical Vaidya spacetime using the Penrose limit geometry around the dynamical photon sphere. In the case of a static spherically symmetric black hole spacetime, it is known that the quasinormal frequency in the eikonal limit can be characterized by the angular velocity and the Lyapunov exponent for the null geodesic congruence on the orbit of the unstable circular null geodesic. This correspondence can be further backed up by the analysis of the Penrose limit geometry around the unstable circular null geodesic orbit. We try to extend this analysis to a Vaidya spacetime, focusing on the dynamical photon sphere in it. Then we discuss to what extent the Penrose limit geometry can be relevant to the ringdown waves in the Vaidya spacetime, comparing the results with the numerically calculated waveform in the Vaidya spacetime. 
\end{abstract}


\pagebreak

\pagestyle{plain}

\section{Introduction}

The detections of the gravitational waves from a binary black hole~\cite{LIGOScientific:2016aoc,LIGOScientific:2017vwq} have opened a new window onto the black holes in our universe. 
The signal from compact object binaries typically consists of three phases: inspiral, merger, and ringdown. 
The ringdown phase, characterized by a superposition of damped sinusoids known as quasinormal modes (QNMs)~\cite{Vishveshwara:1970zz,Press:1971wr}, is particularly significant as it represents the final perturbed black hole settling down to a Kerr black hole~(see, e.g., reviews~\cite{Kokkotas:1999bd,Berti:2009kk,Konoplya:2011qq}). 
The frequencies and damping rates of these QNMs are uniquely determined by the mass and spin of the final Kerr black hole
under the assumption that General Relativity is correct, the effects of surrounding matter fields are negligible, and hence the Kerr hypothesis holds. 
Therefore, measurement of the QNM frequencies offers 
%
a powerful tool for ``black hole spectroscopy", which tests the predictions of General Relativity 
~~\cite{Detweiler:1980gk,Berti:2005ys,Berti:2025hly}. 

Intriguingly, a deep connection has been established between the properties of QNMs in the eikonal (high frequency, geometric optics) limit and the dynamics of null geodesics near the black hole photon sphere (or photon ring) in static or stationary spacetimes~\cite{Mashhoon:1985cya,Ferrari:1984zz,Cardoso:2008bp}. Specifically, the real part of the QNM frequency ($\omega_{\rm Re}$) is related to the angular velocity ($\Omega$) of unstable circular null orbits, while the imaginary part ($\omega_{\rm Im}$), representing the damping rate, is linked to the Lyapunov exponent ($\lambda$) quantifying the instability of these orbits along the radial direction. The correspondence typically takes the form $\omega \approx \Omega \ell - \ii \lambda (n+1/2)$ for large angular momentum quantum number $\ell$ and overtone number $n$. This connection provides valuable physical intuition for the origin of ringdown radiation, associating it with photons (or gravitons) temporarily trapped in unstable orbits near the black hole before escaping to infinity or falling into the black hole. 
This correspondence---referred to as the {\em QNM-geodesic correspondence}--has been checked in various cases of stationary, 
vacuum black holes in 4 and higher dimensions \cite{Cardoso:2008bp}. (See e.g., \cite{Konoplya:2017wot} for the non-vacuum cases, and e.g., \cite{Igata:2025hpy} for further studies.) It is of considerable interest to study whether or not such a correspondence holds in more generic, dynamical cases. 

The Penrose limit (PL)~\cite{Penrose1976} offers a powerful mathematical framework to explore this connection further. By focusing on the geometry in the infinitesimal vicinity of a null geodesic, the PL transforms the spacetime metric into a simpler plane-wave (or pp-wave) form~\cite{Blau_review}. When applied to an unstable circular orbit of photons (UCOP) in a stationary, 
spherically symmetric spacetime, the resulting PL metric inherits the key dynamical properties of the orbit. 
Analyzing wave propagation (e.g., a scalar field) within this simplified geometry naturally reproduces the eikonal QNM spectrum, with the metric components directly related to the orbital frequency $\Omega$ and the Lyapunov exponent $\lambda$, thereby solidifying the QNM-geodesic correspondence~\cite{Blau:2006qj,Cardoso:2008bp,Fransen:2023eqj,Kapec:2024lnr}.
Recent developments of QNM analyses can be seen in, e.g., Refs.~\cite{Giataganas:2024hil,Kehagias:2024sgh,Bucciotti:2025rxa,Kehagias:2025ntm,Perrone:2025zhy,Fransen:2025cgv}. 

However, most astrophysical black holes are not isolated or perfectly stationary; 
they interact with their environment, accreting matter or potentially radiating energy. 
For instance, a black hole formed soon after the neutron star merger would be surrounded by baryonic matter~\cite{LIGOScientific:2017vwq}. 
Furthermore, accretion of dark matter onto a black hole may not be ruled out with electromagnetic observations. 
From a quantum perspective, the black hole will continue to emit Hawking radiation, although this may not be relevant from an astrophysical perspective (see e.g., Ref.~\cite{Xue:2003vs} for calculation of ringdown waveform in an evaporating black hole).
This raises the fundamental question of how QNMs and the ringdown process can be characterized in dynamical spacetimes. 
The Vaidya spacetime~\cite{Vaidya:1951zz}, describing a spherically symmetric radiating or accreting object composed of null dust, provides a simple yet non-trivial model for studying such situations (see Refs.~\cite{Xue:2003vs,Shao:2004ws,Abdalla:2006vb,Bamber:2021knr,Lin:2021fth,Redondo-Yuste:2023ipg,Capuano:2024qhv} for previous works on ringdown waves in dynamical spacetimes). Recent work has shown that the Vaidya spacetime possesses a dynamical photon sphere, whose radius evolves in response to the changing mass function~\cite{Koga:2022dsu}. 

In this paper, we extend the investigation of the QNM-geodesic correspondence to the dynamical context of the Vaidya spacetime. Our primary tool is the PL adapted 
to a null geodesic lying on the dynamical photon sphere trajectory identified in~\cite{Koga:2022dsu}. 
By calculating the PL geometry along this evolving trajectory, we extract time-dependent counterparts of the orbital frequency and Lyapunov exponent within an adiabatic approximation. 
We hypothesize that these quantities characterize the temporal properties of ringdown waves emitted during the dynamical phase. 
To test this hypothesis, we perform numerical simulations of tensor perturbations propagating in the full Vaidya spacetime during both constant and time-dependent accretion phases. 
We then extract the time-varying frequency and damping rate from the resulting ringdown waveform and compare these numerical results with the predictions derived from our PL analysis. This comparison allows us to assess the extent to which the PL geometry around the dynamical photon sphere captures the essential features of ringdown in this non-stationary background.

This paper is organized as follows. In Sec.~\ref{sec:QNMPL}, we review the connection between QNMs and the PL for a static, spherically symmetric spacetime. In Sec.~\ref{sec:linearVaidy}, we analyze the case of a Vaidya spacetime with a constant accretion rate, comparing results from the PL, frequency-domain QNM calculations, and time-domain simulations. In Sec.~\ref{sec:timedepVaidya}, we extend this analysis to Vaidya spacetimes with time-dependent accretion rates, comparing the predictions from the adiabatic PL with full numerical ringdown simulations. Section~\ref{sec:summary_conclusions} provides a summary and discussion of our findings. We use geometric units where $G=c=1$ throughout the paper.
 
\section{Quasinormal modes in the Penrose limit of a static spherically symmetric spacetime}
\label{sec:QNMPL}

\subsection{Unstable circular orbits of photons}
Since we are interested in the PL geometry adapted to a UCOP, let us review the null geodesics and UCOP in a static spherically symmetric spacetime in this subsection. 
Let us write the metric of a static spherically symmetric spacetime as follows:
\begin{equation}
  \dd s^2=-f(r)\dd t^2+h(r)\dd r^2+r^2\dd \theta^2+r^2 \sin^2 \theta \dd \phi^2. 
\end{equation}
Because of the symmetry, we can focus on the equatorial surface, and in what follows we set $\theta=\pi/2$ without loss of generality. 
Then the Lagrangian for 
a free particle motion 
is given by 
\begin{equation}
  \mathcal L=\frac{1}{2}\left(-f\dot t^2+h\dot r^2+r^2\dot \phi^2\right) , 
\end{equation}
where the dot `` $\dot{}$ " denotes the derivative with respect to an affine parameter and ${\cal L}$ is set to zero for a null geodesic.
The equations of motion 
are obtained as 
\begin{eqnarray}
  \dot t&=&\frac{E}{f}, \\
  \dot \phi&=&\frac{L}{r^2}, \\
  0&=&\frac{1}{2}\dot r^2+V(r), 
\end{eqnarray}
where $E$ and $L$ are the constants of motion associated with the time translation and rotation symmetries, 
and the effective potential $V(r)$ is given as 
\begin{equation}
  V(r)=-\frac{\Delta^2}{2fhr^2}
\end{equation}
with 
\begin{equation}
  \Delta=\sqrt{E^2r^2-fL^2}. 
\end{equation}
The second-order derivative for the radial motion can be obtained as 
\begin{equation}
  \ddot r=-V'(r), 
  \label{eq:ddr}
\end{equation}
where the prime `` ${}'$ " describes the derivative with respect to $r$. 

For the realization of UCOP given by $r=r_\rmc$, 
$E$, $L$ and $r_\rmc$ must satisfy
\begin{equation}
  V(r_\rmc)=V'(r_\rmc)=0. 
\end{equation}
These two conditions can be rewritten as 
\begin{eqnarray}
  0&=&\Delta_\rmc^2:=E^2r_\rmc^2-f_\rmc L^2, \\
  0&=&2f_\rmc-r_\rmc f'_\rmc, 
\end{eqnarray}
where $f_\rmc=f(r_\rmc)$ and $f'_\rmc=f'(r_\rmc)$. 
Then we obtain
\begin{equation}
  V''(r_\rmc)=\frac{L^2}{2h_\rmc r_\rmc^2}\left(-\frac{2}{r_\rmc^2}+\frac{f''_\rmc}{f_\rmc}\right), 
\end{equation}
where $f''_\rmc=f''(r_\rmc)$. 

This circular orbit is unstable against small deviations along the radial direction. 
The unstable behavior can be characterized by the characteristic exponent $\lambda$ as $\sim \ee^{\lambda t}$. 
Substituting the functional form of the small deviation $\delta r\sim \ee^{\lambda t}$ into Eq.~\eqref{eq:ddr}, 
we obtain 
\begin{equation}
  \lambda^2=-\frac{V''_\rmc}{\dot t|_{r=r_\rmc}^2}=-V''_\rmc\frac{f_\rmc^2}{E^2}. 
\end{equation}
On the other hand, for the independent small deviation to the $\theta$-direction with $r=r_\rmc$ and $\dot r_\rmc=0$, we obtain the circular motion which intersects with the original circular trajectory at two antipodal points. 
Therefore, the circular orbit is stable against such a deviation, and the frequency for the stable oscillatory behavior is given by the angular frequency for the circular motion, that is,
\begin{equation}
  \Omega:=\left.\frac{\dd}{\dd t}\phi\right|_{r=r_\rmc}=\left.\frac{\dot \phi}{\dot t}\right|_{r=r_\rmc}=\frac{L/r_\rmc^2}{E/f_\rmc}=\frac{\sqrt{f_\rmc}}{r_\rmc}. 
\end{equation}
For the Schwarzschild black hole, we obtain $\lambda=\Omega=1/(3\sqrt{3}M_{\rm Sch})$ with $M_{\rm Sch}$ being the mass of the Schwarzschild black hole. 

\subsection{Coordinate transformation and Penrose limit}

Here we derive the PL metric in an ad hoc way, for the sake of a simple argument. 
The general derivation of the PL is shown in Appendix~\ref{sec:PL}, and the derivation of the Penrose coordinates adapted to a UCOP is described in Appendix~\ref{sec:PLPS}. 
First, let us express a UCOP in a parametric form by using the affine parameter $\tau$ as 
\begin{eqnarray}
  t&=&\frac{E}{f_\rmc}\tau, \\
  r&=&r_\rmc, \\
  \theta&=&\frac{\pi}{2}, \\
  \phi&=&\frac{L}{r_\rmc^2}\tau=\frac{E}{r_\rmc \sqrt{f_\rmc}}\tau=:\beta \tau. 
\end{eqnarray}
Then we introduce the following new coordinates $(u,v,x,y)$:
\begin{eqnarray}
  t&=&\frac{E}{f_\rmc}(u+\varepsilon^2\Lambda v)-\frac{\varepsilon^2}{E}v, 
  \label{eq:tuv0}\\
  r&=&r_\rmc +\frac{\varepsilon}{\sqrt{h_\rmc}}x, \\
  \theta&=&\frac{\pi}{2}+\frac{\varepsilon}{r_\rmc}y, \\
\phi&=&\beta (u+\varepsilon^2\Lambda v), 
\label{eq:phuv}
\end{eqnarray}
where $\varepsilon$ corresponds to the scaling parameter for taking the PL, 
and $\Lambda$ is an arbitrary constant corresponding to a gauge degree of freedom (see Refs.~\cite{Fransen:2023eqj,Kapec:2024lnr}, and Eqs~\eqref{eq:tuv} and \eqref{eq:thuv}). 
We note that $\dd x=\sqrt{h_\rmc} \dd r$ and $\dd y=r_\rmc \dd \theta$ compose zweibeins\footnote{$\dd x=\sqrt{h_\rmc} \dd r$ is not parallelly transported along the null geodesic differently from $E^i_{(a)}$ defined in Appendix~\ref{sec:PL}. }, and 
$\del_u$ is the null geodesic generator.
Then we obtain the desired metric in a similar way as in Eq.~\eqref{eq:PLlimit} as follows:
\begin{eqnarray}
  \dd \bar s^2&=&\lim_{\varepsilon\rightarrow0}\left(\varepsilon^{-2}\left.\dd s^2\right|_{(t,r,\theta,\phi)\rightarrow(u,v,x,y)}\right), \\
  &=&\dd u\dd v+(\alpha^2 x^2-\beta^2 y^2)\dd u^2+\dd x^2+\dd y^2, 
   \label{eq:PLlimit_sp}
\end{eqnarray}
where $\alpha=-V''_\rmc$. 
This is the PL metric adapted to a UCOP in the Brinkmann coordinates. 
The coefficients $\alpha$ and $\beta$ are related to $\Omega$ and $\lambda$ as 
\begin{eqnarray}
\alpha&=&\dot t|_{r=r_\rmc}\lambda=\frac{E}{f_\rmc}\lambda, 
\label{eq:alphalambda}\\
\beta&=&\dot t|_{r=r_\rmc}\Omega=\frac{E}{f_\rmc}\Omega.
\label{eq:betaomega}
\end{eqnarray}
The value of $E$ can be set to 1 for the static case by performing the constant scaling 
of the affine parameter $\tau$. 
It should be noted that, for the dynamical case, which we consider in the main text, the value of $E=f_\rmc \dot t$ is not a constant of motion anymore.

\subsection{Quasinormal modes in the PL}

Let us consider a massless scalar field $\Psi$ in the PL metric \eqref{eq:PLlimit_sp}.
In this section, we basically follow the discussion in Ref.~\cite{Kapec:2024lnr}. 
The wave equation is given by 
\begin{equation}
  \nabla_\mu\nabla^\mu\Psi=\left[2\del_u\del_v-\left(\alpha^2x^2-\beta^2y^2\right)\del_v^2+\del_x^2+\del_y^2\right]\Psi=0. 
\end{equation}
Assuming the form of the separation of variables
\begin{eqnarray}
  \Psi=\ee^{-\ii \tilde \omega u+\ii p v}X(x)Y(y), 
\end{eqnarray}
we obtain 
\begin{eqnarray}
  \mu_xX&=&-\left(\frac{\dd^2}{\dd x^2} +p^2\alpha^2x^2\right)X, 
  \label{eq:IHO}\\
  \mu_yY&=&-\left(\frac{\dd^2}{\dd y^2} -p^2\beta^2y^2\right)Y, 
  \label{eq:SHO}\\
  0&=&2\tilde \omega p-\mu_x-\mu_y. 
\end{eqnarray}
The first/second equation is the inverted/standard harmonic oscillator (IHO/SHO). 
The eigenvalues for the SHO is simply given by 
\begin{equation}
  \mu_y=2\beta p \left(n_y+\frac{1}{2}\right), 
\end{equation}
where $n_y$ takes a non-negative integer. 
For the IHO, we are interested in the pure outgoing solutions, which can be 
obtained just replacing the parameter $p$ by $-\ii p$,%
\footnote{The IHO equation can be obtained by $p\rightarrow \pm \ii p$. 
Then, including the time dependence, the asymptotic form of the eigen function transforms as $\exp\left[-\ii \tilde \omega u - \alpha p x^2/2\right]\rightarrow \exp\left[-\ii(\tilde \omega u \pm \alpha p x^2/2)\right]$. Therefore the lower sign corresponds to the pure outgoing boundary condition. } 
namely, 
\begin{equation}
  \mu_x=-2\ii \alpha p\left(n_x+\frac{1}{2}\right) 
\end{equation}
with $n_x$ being a non-negative integer. 
Then we obtain 
\begin{equation}
  \tilde \omega=\beta\left(n_y+\frac{1}{2}\right)-\ii \alpha\left(n_x+\frac{1}{2}\right). 
\end{equation}

In the standard way of the separation of variables with respect to the original coordinates $(t,r,\theta,\phi)$, 
the wave function is proportional to $\ee^{-\ii \omega t+\ii m\phi}$ with the frequency $\omega$ 
and a magnetic quantum number $m$. 
Rewriting this factor in terms of the coordinates $u$ and $v$ through Eqs.~\eqref{eq:tuv0} and \eqref{eq:phuv}, we obtain
\begin{equation}
  \exp\left[-\ii \omega t+\ii m \phi\right]=\exp\left[\ii \left(\frac{\omega}{E}-\frac{E\omega}{f_\rmc}\Lambda+m\beta\Lambda\right)v-\ii\left(\frac{L^2}{Er_\rmc^2}\omega-\beta m\right)u\right], 
\end{equation}
where we have reset the bookkeeping parameter $\varepsilon$ to $1$, or in other words, renormalized the coordinates $v$ including $\varepsilon^2$. 
Hereafter, we also renormalize the coordinates $x$ and $y$ by simply setting  $\varepsilon=1$. 
Since the exponent $-\ii \omega t+\ii m \phi$ must be equal to $-\ii \tilde \omega u+\ii pv$ in the PL, we obtain the following relations:
\begin{eqnarray}
  p&=&\frac{\omega}{E}-\frac{E\omega}{f_\rmc}\Lambda+m\beta\Lambda, \\
  \tilde \omega&=&\beta\left(\frac{L}{E}\omega-m\right). 
\end{eqnarray}
The second equation can be further rewritten as 
\begin{eqnarray}
  \omega&=&\frac{E}{L}\left(n_y+m+\frac{1}{2}\right)-\ii\frac{E}{L}\frac{\alpha}{\beta}\left(n_x+\frac{1}{2}\right)\cr
  &=&\Omega\left(n_y+m+\frac{1}{2}\right)-\ii\lambda \left(n_x+\frac{1}{2}\right). 
  \label{eq:corresp}
\end{eqnarray}
Therefore, identifying $n_y+m$ to $\ell$ we can establish the correspondence between the null geodesics and QNMs reported in Ref.~\cite{Cardoso:2008bp}. 
Since $\omega$ is a complex variable, 
$p$ is also complex in general, and the amplitude of $\Psi$ changes along the $v$-direction unless $p$ is a real value. 
To make the value of $p$ real, 
we need to fix the 
arbitrary constant $\Lambda$ to be $\frac{f_\rmc}{E^2}$. 
Then we obtain $p=m/L$, and the relevant components of the coordinate transformation become
\begin{eqnarray}
  t&=&\frac{E}{f_\rmc}u, \\
\phi&=&\beta u+\frac{v}{L}.  
\end{eqnarray}
We find $\del_\phi\propto\del_v$ in this case, and there is no decay of the amplitude along the $v$-direction. 

We can further justify this correspondence by considering the waveform along $y$ or equivalently $\theta$ direction. 
First, it should be noted that, since we are interested in the vicinity of a circular orbit, this correspondence is expected to be correct for the case $\ell>|m|\gg1$. 
Let us focus on the associated Legendre function in the spherical harmonics with the magnetic quantum number $m$ 
being the highest value $m=\ell$:
\begin{equation}
  P^\ell_\ell(\cos\theta)\sim\left(1-\cos^2\left(\frac{\pi}{2}+\frac{y}{r_\rmc}\right)\right)^{\ell/2}, 
\end{equation}
where we omitted the constant factor for simplicity. 
Taking the Taylor expansion with respect to $y/r_\rmc$, we obtain 
\begin{equation}
  P^\ell_\ell(\cos\theta)\sim 1-\frac{\ell}{2}\frac{y^2}{r_\rmc^2}. 
\end{equation}
A similar expression can be obtained by expanding the ground state for the SHO described by Eq.~\eqref{eq:SHO} as follows:
\begin{equation}
  \exp\left[-\frac{1}{2}\beta p y^2\right]
  \sim 
  \exp\left[-\frac{1}{2}\beta \frac{m}{L} y^2\right]
  =\exp\left[-\frac{1}{2}\frac{mE}{f_\rmc L} \Omega y^2\right]\sim 1-\frac{n_y+m}{2}\frac{y^2}{r_\rmc^2}, 
\end{equation}
where $n_y=0$ for the ground state. 
This expression also suggests the relation $\ell=n_y+m$. 
Moreover, we can rewrite the lowering operator in the spherical harmonics $\propto (-\del_\theta+\ii\cot\theta\del_\phi)$ as follows:
\begin{equation}
  -\del_\theta+\ii\cot\theta\del_\phi
  \sim
  -r_\rmc \del_y- \cot\left(\frac{y}{r_\rmc}+\frac{\pi}{2}\right)m
  \sim -r_\rmc \del_y+\frac{L}{r_\rmc}py=r_\rmc(-\del_y+\beta p y). 
\end{equation}
This is proportional to the raising operator for SHO. 
This fact also gives a support to the relation $\ell=n_y+m$. 
Explicit comparison of the waveforms can be seen in Ref.~\cite{Kapec:2024lnr}. 
Since we are interested in the regime in which the PL approximation works at a certain extent, 
we focus on the cases $\ell\geq4$ in this paper.
Although $\ell=4$ may not appear to be a large enough value to justify the eikonal approximation, we will demonstrate that even the $\ell=4$ case provides very accurate predictions, as has already been pointed out by the existing literature~\cite{Iyer:1986nq,Berti:2005eb}. 
We will elaborate on the cases $\ell<4$ in a future study focusing on the behavior of the power-law tail.

\section{Penrose limit and Ringdown analysis in a Vaidya spacetime with a constant accretion rate}
\label{sec:linearVaidy}
So far, we have been considering static spherically symmetric spacetimes. 
From this section, we treat the Vaidya spacetime described by 
\begin{equation}
  \dd s^2=-\left(1-\frac{2M(\mathcal V)}{r}\right)\dd \mathcal V^2+2\dd \mathcal V\dd r+r^2\left(\dd \theta^2+\sin^2\theta\dd \phi^2\right). 
  \label{eq:metVaidya}
\end{equation}
First, we focus on the mass function linearly increasing with the time coordinate $\mathcal V$:
\begin{equation}
  M(\mathcal V)=M_1+M'\mathcal V
\end{equation}
with $M'={\rm const.}>0$. 
The frequency-domain analyses for the static conformal metric have been done in Ref.~\cite{Capuano:2024qhv}. 
We reanalyze it in the time domain and compare the results with the frequency domain analyses. 

Following Ref.~\cite{Capuano:2024qhv}, let us introduce the following new coordinates:
\begin{eqnarray}
x&=&\frac{r}{2M(\mathcal V)}, \\ 
T&=&\int\frac{\dd \mathcal V}{2M(\mathcal V)}-x_*(x) \\
&=&\frac{1}{2M'}\ln\left|1+M'\frac{\mathcal V}{M_1}\right|-x_*(x),  
\end{eqnarray}
where $x_*$ is the generalized tortoise coordinate
\begin{equation}
  x_*(x)=\int\frac{\dd x}{\tilde f(x)}
\end{equation}
with 
\begin{equation}
  \tilde f(x)=1-\frac{1}{x}-4M'x. 
\end{equation}
Then the metric form can be rewritten as 
\begin{equation}
  \dd s^2=g^{(0)}_{\mu\nu}\dd \tilde x^\mu \dd \tilde x^\nu:=4M^2(\mathcal V)\left[-\tilde f(x)\dd T^2+\frac{1}{\tilde f(x)}\dd x^2+x^2\dd\Omega^2\right],  
\end{equation}
where $\tilde x^\mu$ denotes the coordinates $(T, x, \theta, \phi)$. 
The background metric $g^{(0)}_{\mu\nu}$ is conformal to the static metric $\tilde g^{(0)}_{\mu\nu}$ given by 
\begin{equation}
  \tilde g^{(0)}_{\mu\nu}\dd \tilde x^\mu \dd \tilde x^\nu=-\tilde f(x)\dd T^2+\frac{1}{\tilde f(x)}\dd x^2+x^2\dd\Omega^2. 
\end{equation}
Two horizons are located at the radius satisfying $\tilde g^{(0)xx}=\tilde f(x)=0$, namely, 
\begin{equation}
  x=x_\mp:=\frac{1\mp\sqrt{1-16M'}}{8M'}. 
\end{equation}
The horizon $x=x_-$ corresponds to the black hole event horizon, and $x=x_+$ indicates the external cosmological horizon. 
In this paper, we consider $M'<1/16$ and, in the following, restrict our attention to the conformally static region $x_-<x<x_+$.

\subsection{Penrose limit analysis}

\begin{figure}[h!]
\begin{center}
\includegraphics[width=12cm]{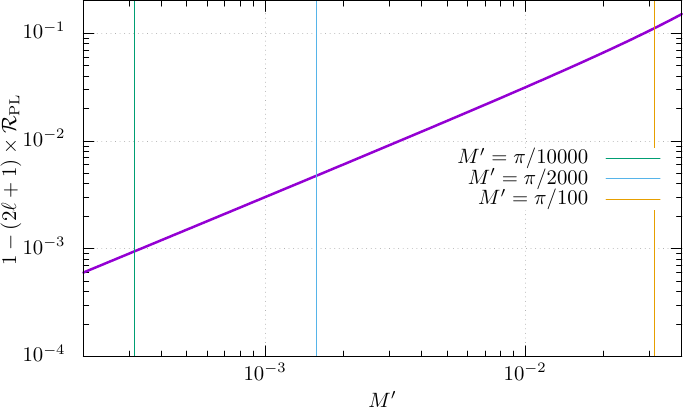}
\caption{\baselineskip5mm
The value of $1-(2\ell+1)\mathcal R_{\rm PL}$ is shown as a function of $M'$. 
The specific values of $M'$ indicated by the vertical lines correspond to the values of $M'$ (Sec.~III) and 
$M'_{\rm max}$ (Sec.~IV) for specific examples shown in the subsequent sections. 
}
\label{fig:RPL}
\end{center}
\end{figure}

Since the null geodesic trajectories are invariant under the conformal transformation, let us focus on the conformal static metric $\tilde g^{(0)}_{\mu\nu}$. 
The radius for the circular photon orbit $x_{\rm c}$ can be given by the equation
$2\tilde f(x_{\rm c})-x_{\rm c}\tilde f'(x_{\rm c})=0$ as 
\begin{equation}
  x_{\rm c}=\frac{1-\sqrt{1-12M'}}{4M'}, 
\end{equation}
where we took the smaller root associated with the event horizon. 
Then the QNM frequency of the fundamental mode $n=0$ can be estimated as 
\begin{eqnarray}
  \Omega^{\rm Re}_{\rm PL}&=&\frac{\sqrt{\tilde f(x_{\rm c})}}{x_c}\left(\ell+\frac{1}{2}\right), \\
  \Omega^{\rm Im}_{\rm PL}&=&-\frac{1}{2}\sqrt{\frac{\tilde f(x_\rmc)^2}{x_\rmc^2}-\frac{1}{2}\tilde f(x_\rmc) \tilde f''(x_\rmc)}. 
\end{eqnarray}
More specifically, we obtain
\begin{eqnarray}
  \Omega^{\rm Re}_{\rm PL}&=&
  -\frac{2 \sqrt{6 \sqrt{1-12 M'}-3} }{3 \left(\sqrt{1-12 M'}-1\right)}M' (2 \ell+1), \\
  \Omega^{\rm Im}_{\rm PL}&=&-\frac{\sqrt{6 M' \left(24 M'-3 \sqrt{1-12 M'}-4\right)+\sqrt{1-12 M'}+1}}{3 \sqrt{6}}.   
\end{eqnarray}
For $M'=\pi/2000$ and $\ell=4$, we obtain $\Omega^{\rm Re}_{\rm PL}=1.70745$ and $\Omega^{\rm Im}_{\rm PL}=-0.188816$. 
The ratio $\mathcal R_{\rm PL}$ between $\Omega^{\rm Im}_{\rm PL}$ and $\Omega^{\rm Re}_{\rm PL}$ is given by 
\begin{equation}
  \mathcal R_{\rm PL}:=\left|\frac{\Omega^{\rm Im}_{\rm PL}}{\Omega^{\rm Re}_{\rm PL}}\right|=
  \frac{\left(\sqrt{1-12 M'}-1\right) \sqrt{6 M' \left(24 M'-3 \sqrt{1-12 M'}-4\right)+\sqrt{1-12 M'}+1}}{6 (2\ell+1)M'\sqrt{4 \sqrt{1-12 M'}-2} }. 
\end{equation}
The value of $1-(2\ell+1)\mathcal R_{\rm PL}$ is shown as a function of $M'$ in Fig.~\ref{fig:RPL}. 
In Fig.~\ref{fig:RPL}, the specific values of $M'$ indicated by the vertical lines correspond to the values of $M'$ (Sec.~III) and 
$M'_{\rm max}$ (Sec.~IV) for specific examples shown in the subsequent sections. 
The ratio $(2\ell+1)\mathcal R_{\rm PL}$ decreases with the value of $M'$ compared to 1 for the Schwarzschild case. 
\subsection{Frequency domain analyses}

Let us consider the massless scalar field, odd-parity Maxwell field, and odd-parity metric perturbations. 
For the scalar field $\Phi$, we consider the conformal transformation with the conformal weight $-1$, namely, we introduce the scalar field $\tilde \Phi$ as follows:
\begin{equation}
  \Phi=(2M(\mathcal V))^{-1}\tilde \Phi. 
\end{equation}
The scalar field equation of motion is given by \cite{Wald:1984rg}
\begin{equation}
  \nabla_\mu\nabla^\mu \Phi=\left(\tilde \nabla_\mu \tilde \nabla^\mu -\frac{\tilde R}{6}\right)\tilde \Phi=0, 
\end{equation}
where $\tilde R$ is the Ricci scalar for the conformal metric $\tilde g_{\mu\nu}$. 
Then, substituting the following form of the mode decomposition into the equation of motion: 
\begin{equation}
  \tilde \Phi=\ee^{-\ii \tilde \Omega T}\frac{\psi^{(0)}(x)}{x}Y_{\ell m}(\theta,\phi), 
\end{equation}
we obtain
\begin{equation}
  \left[\frac{\dd^2}{\dd x_*^2}+\tilde \Omega^2-\tilde f(x)\left(\frac{\ell(\ell+1)}{x^2}+\frac{1}{x^3}\right)\right]\psi^{(0)}(x)=0. 
\end{equation}

For the Maxwell field, we introduce the odd parity perturbation as follows:
\begin{equation}
  A_\mu\dd \tilde x^\mu=\ee^{-\ii \tilde \Omega T}\psi^{(1)}(x)\epsilon_i^{~j}\del_j Y_{\ell m} \dd \tilde x^i, 
\end{equation}
where $i$ and $j$ run over indices corresponding to the spherical coordinates $\theta$ and $\phi$, and $\epsilon_{ij}$ is the totally anti-symmetric tensor on the two-sphere. 
Then, from the Maxwell equation, we obtain 
\begin{equation}
  \left[\frac{\dd^2}{\dd x_*^2}+\tilde \Omega^2-\tilde f(x)\left(\frac{\ell(\ell+1)}{x^2}\right)\right]\psi^{(1)}(x)=0. 
\end{equation}

For the odd parity gravitational wave mode, in the Regge-Wheeler gauge, the metric perturbation $\tilde h_{\mu\nu}$ for the conformal metric $\tilde g_{\mu\nu}=\tilde g^{(0)}_{\mu\nu}+\tilde h_{\mu\nu}$ can be introduced as 
\begin{eqnarray}
  \tilde h_{\mu\nu}\dd \tilde x^\mu \dd \tilde x^\nu&=&\ee^{-\ii \tilde \Omega T}\Biggl[2h_0(x)\left(-\frac{1}{\sin\theta}\del_\phi Y_{\ell m}\dd T\dd \theta+\sin\theta \del_\theta Y_{\ell m}\dd T\dd\phi\right)\cr
  &&+2h_1(x)\left(-\frac{1}{\sin\theta}\del_\phi Y_{\ell m}\dd x\dd \theta+\sin\theta \del_\theta Y_{\ell m}\dd x\dd\phi\right)
  \Biggr]. 
  \label{eq:tilh}
\end{eqnarray}
In terms of the conformal metric $\tilde g_{\mu\nu}$, the gravitational field equations are given by \cite{Wald:1984rg}
\begin{equation}
  \tilde R_{\mu\nu}-2\tilde \nabla_\mu\tilde \nabla_\nu \ln M -\tilde g_{\mu\nu}\tilde \nabla_\alpha\tilde \nabla^\alpha \ln M+2\tilde \nabla_\mu\ln M\tilde \nabla_\nu \ln M-2\tilde g_{\mu\nu}\tilde \nabla_\alpha\ln M \tilde \nabla^\alpha \ln M=8\pi T_{\mu\nu}
\end{equation}
with 
\begin{equation}
  T_{\mu\nu}\dd \tilde x^\mu\dd \tilde x^\nu=
  \frac{M'}{4\pi x^2}\left[\dd T^2+\frac{2}{\tilde f(x)}\dd T\dd x+\frac{\dd x^2}{\tilde f(x)^2}\right].  
\end{equation}
We note that $\tilde{\nabla}_\mu \ln M$ does not depend on $T$ 
in the coordinates $\tilde{x}^\mu$ since $(\tilde{\nabla}_\mu \ln M )\dd\tilde{x}^\mu = 2 M^\prime(\dd T + \dd x/\tilde{f})$.
Introducing the master variable $\psi^{(2)}(x):=F(x)^{-1}h_1(x)$ with 
\begin{equation}
  F(x)=-\frac{x^{5/4}}{f(x)^{3/4}}\exp\left(\frac{{\rm arctan}\left(\frac{8M'x-1}{\sqrt{16M'-1}}\right)}{2\sqrt{16M'-1}}\right), 
\end{equation}
one can obtain the following master equation:
\begin{equation}
  \left[\frac{\dd^2}{\dd x_*^2}+\left(\tilde \Omega+2\ii M'\right)^2-\tilde f(x)\left(\frac{\ell(\ell+1)}{x^2}-\frac{3}{x^3}\right)\right]\psi^{(2)}(x)=0.   
\end{equation}
In summary, the master equations can be written as 
\begin{equation}
  \left[\frac{\dd^2}{\dd x_*^2}+\hat \Omega^2-\tilde f(x)\left(\frac{\ell(\ell+1)}{x^2}-\frac{s^2-1}{x^3}\right)\right]\psi^{(s)}(x)=0,      
\end{equation}
where $\hat \Omega=\tilde \Omega$ for $s=0$ and $s=1$, and $\hat \Omega=\tilde \Omega+2\ii M'$ for $s=2$. 
The master equations can be solved with purely ingoing/outgoing boundary conditions at $x=x_\mp$. 
The results obtained by the Leaver's method \cite{Leaver:1985ax} (see, e.g., a review \cite{Konoplya:2020bxa}) 
with $M'=\pi/2000$ and $\ell=4$ are summarized in Table \ref{tab:frequencies}. 
\begin{table}[h!]
\caption{The real and imaginary QNM frequencies in the conformal static spacetime of the metric $\tilde g_{\mu\nu}$ with $M'=\pi/2000$ and $\ell=4$. }
\label{tab:frequencies}
\begin{tabularx}{165mm}{cc|C|C|C}
 \hline
Field&Master variable~~&${\rm Re}\hat \Omega$ &${\rm Im}\hat \Omega$&$|{\rm Im}\hat \Omega/{\rm Re}\hat \Omega|$\\
\hhline{==|=|=|=}
PL&---&$1.70745$ & $-0.188816$ &$0.110584$\\
\hline
Massless scalar&$\psi^{(0)}(x)$&$1.71037$ & $-0.189146$ & $0.110588$ \\
\hline
Electromagnetic&$\psi^{(1)}(x)$&$1.68210$ &$-0.188106$ & $0.111828$\\
\hline
Metric perturbation&$\psi^{(2)}(x)$&$1.59536$ & $-0.184788$& $0.115828$\\
\hline
\end{tabularx}
\end{table}

\subsection{Time-domain analyses}
\label{sec:timedomain_linear}

To compare the analysis of the PL geometry with the ringdown waveform in Vaidya spacetimes, we numerically solve the tensor mode master equation in the time domain by using 
the double null formalism following Refs.~\cite{Burko:1997tb,Eilon:2015axa}. 
First, let us consider the relation between the double null coordinates $(\mathcal U,\mathcal V$) and the coordinates $(\mathcal V, r)$ in Eq.~\eqref{eq:metVaidya}. 
In the double null coordinates, the metric form is given as 
\begin{equation}
  \dd s^2=-\ee^{\sigma(\mathcal U,\mathcal V)}\dd\mathcal U\dd \mathcal V+r^2(\mathcal U,\mathcal V)(\dd \theta^2+\sin^2\theta\dd\phi^2). 
\end{equation}
Assuming the coordinate transformation of the form 
\begin{equation}
  \dd\mathcal U=A(\mathcal V,r)\dd \mathcal V-B(\mathcal V,r)\dd r, 
\end{equation}
from the null condition $g^{\mu\nu}(\dd \mathcal U)_\mu (\dd \mathcal U)_\nu=0$, we obtain
\begin{equation}
  A=\frac{1}{2}\left(1-\frac{2M(\mathcal V)}{r}\right)B. 
\end{equation}
Then we find 
\begin{equation}
  \dd r=-\frac{1}{B}\dd \mathcal U+\frac{1}{2}\left(1-\frac{2M(\mathcal V)}{r}\right)\dd \mathcal V,
\end{equation}
so that 
\begin{eqnarray}
  \frac{\del r}{\del \mathcal U}&=&-\frac{1}{B}=-\frac{1}{2}\ee^{\sigma(\mathcal U,\mathcal V)}, \\
  \frac{\del r}{\del \mathcal V}&=&\frac{1}{2}\left(1-\frac{2M(\mathcal V)}{r}\right), 
  \label{eq:rV}\\
  \frac{\del^2r}{\del\mathcal U\del\mathcal V}&=&\frac{\del}{\del \mathcal U}\left[\frac{1}{2}\left(1-\frac{2M(\mathcal V)}{r}\right)\right]. 
  \label{eq:rUV}
\end{eqnarray}
We numerically solve Eq.~\eqref{eq:rUV} with the boundary condition Eq.~\eqref{eq:rV} on the 
outermost grid points along $\mathcal U={\rm const.}$ line. 

For the massless scalar field $\Phi$, let us introduce a new variable $\Psi^{(0)}(\mathcal U,\mathcal V)$ as follows:
\begin{equation}
  \Phi=\frac{\Psi^{(0)}(\mathcal U,\mathcal V)}{r(\mathcal U,\mathcal V)}Y_{\ell m}(\theta,\phi). 
\end{equation}
Then, the massless scalar equation of motion is given by 
\begin{equation}
  \frac{\del^2 \Psi^{(0)}}{\del\mathcal U\del\mathcal V}=-\frac{1}{2r^2}\frac{\del r}{\del\mathcal U}\left[\ell(\ell+1)+\frac{2M(\mathcal V)}{r}\right]\Psi^{(0)}. 
\end{equation}

For the Maxwell field, we introduce the master variable $\Psi^{(1)}$ defined by 
\begin{equation}
  A_\mu\dd x^\mu=\Psi^{(1)}(\mathcal U,\mathcal V)\epsilon_i^{~j}\del_j Y_{\ell m} \dd x^i, 
\end{equation}
where $x^\mu$ denotes the coordinates $(\mathcal U, \mathcal V, \theta, \phi)$. 
Then, from the Maxwell equation, we obtain 
\begin{equation}
  \frac{\del^2 \Psi^{(1)}}{\del\mathcal U\del\mathcal V}=-\frac{1}{2r^2}\frac{\del r}{\del\mathcal U}\left[\ell(\ell+1)\right]\Psi^{(1)}. 
\end{equation}

For the odd parity metric perturbation, in the Regge-Wheeler gauge, the metric perturbation can be described as 
\begin{eqnarray}
  h_{\mu\nu}\dd x^\mu \dd x^\nu&=&
  2H_0(\mathcal U,\mathcal V)\left(-\frac{1}{\sin\theta}\del_\phi Y_{\ell m}\dd \mathcal U\dd \theta+\sin\theta \del_\theta Y_{\ell m}\dd \mathcal U\dd\phi\right)\cr
  &&+2 H_1(\mathcal U,\mathcal V)\left(-\frac{1}{\sin\theta}\del_\phi Y_{\ell m}\dd \mathcal V\dd \theta+\sin\theta \del_\theta Y_{\ell m}\dd \mathcal V\dd\phi\right).
  \label{eq:hH}
\end{eqnarray}
Introducing the dimensionless master variable $\Psi^{(2)}(\mathcal U,\mathcal V)$ given by
\begin{equation}
  \Psi^{(2)}(\mathcal U,\mathcal V)=-\frac{4\ee^{-\sigma}}{(\ell-1)(\ell+2)M_1}\left[2\left(\frac{\del r}{\del \mathcal V} H_0-\frac{\del r}{\del \mathcal U}  H_1\right)+r\left(\frac{\del H_1}{\del \mathcal U}-\frac{\del H_0}{\del \mathcal V}\right)\right], 
\end{equation}
we obtain the following master equation~\cite{Gerlach:1979rw}
\begin{equation}
  \frac{\del^2 \Psi^{(2)}}{\del \mathcal U \del\mathcal V }=-\frac{1}{2r^2}\frac{\del r}{\del \mathcal U}\left[\ell(\ell+1)-\frac{6M(\mathcal V)}{r}\right]\Psi^{(2)}. 
\end{equation}
The metric components $H_0$ and $H_1$ can be calculated from the master variable $\Psi^{(2)}$ as follows: 
\begin{eqnarray}
  H_0&=& \frac{M_1}{2}\left(\frac{\del r}{\del \mathcal U}\Psi^{(2)}+r\frac{\del\Psi^{(2)}}{\del\mathcal U}\right),
  \label{eq:H0}\\
  H_1&=& \frac{M_1}{2}\left(\frac{\del r}{\del \mathcal V}\Psi^{(2)}+r\frac{\del\Psi^{(2)}}{\del\mathcal V}\right).
  \label{eq:H1} 
\end{eqnarray}

In summary, we obtain the master equations
\begin{equation}
  \frac{\del^2 \Psi^{(s)}}{\del \mathcal U \del\mathcal V }=-\frac{1}{2r^2}\frac{\del r}{\del \mathcal U}\left[\ell(\ell+1)-(s^2-1) \frac{2M(\mathcal V)}{r}\right]\Psi^{(s)}
  \label{eq:master}
\end{equation}
for the massless scalar ($s=0$), Maxwell ($s=1$), and metric perturbations ($s=2$). 

For the Schwarzschild spacetime, 
with the mass $M$, the master equation \eqref{eq:master} can be reduced to the form
\begin{equation}
  \left[\frac{\dd^2}{\dd r_*^2}+\omega_{\rm Sch}^2-\left(1-\frac{2M}{r}\right)\left(\frac{\ell(\ell+1)}{r^2}-\frac{s^2-1}{r^3}\right)\right]\psi_{\rm Sch}^{(s)}(x)=0,      
\end{equation}
  where $r_*$ is the tortoise coordinate, and we have introduced the frequency $\omega_{\rm Sch}$. 
  For later convenience, we show the values of $\omega^{\rm Im/Re}_{\rm PL}$ and 
numerically calculated values of  $\omega_{\rm Sch}^{\rm Im/Re}$ of the fundamental mode in the Schwarzschild spacetime 
with $\ell=4$ in Table~\ref{tab:PLSch}. 
We also calculate the numerical factor $\mathcal C:=(2\ell+1)|\omega^{\rm Im}/\omega^{\rm Re}|$, 
which gives $1$ for the Penrose-limit analysis. 
The values of $\mathcal C$ can be regarded as correction factors for the value of $(2\ell+1)|\omega^{\rm Im}/\omega^{\rm Re}|$ of each field 
needed for comparing the PL analysis with the numerical ringdown analyses.  
It should be noted that we focus on the fundamental modes in this paper, and the values of $\mathcal C$ are dependent on the overtone number. 
\begin{table}[h!]
\caption{The values of $\omega_{\rm Sch}^{\rm Im/Re}$ and $\mathcal C=(2\ell+1)|\omega_{\rm Sch}^{\rm Im}/\omega_{\rm Sch}^{\rm Re}|$ of the fundamental mode in the Schwarzschild spacetime 
with $\ell=4$ with the unit $M=1$. }
\label{tab:PLSch}
\begin{tabularx}{165mm}{c|C|C|C}
\hline
&$\omega_{\rm Sch}^{\rm Re}$&$\omega_{\rm Sch}^{\rm Im}$&$\mathcal C$
\\
\hhline{=|=|=|=}
PL&$\frac{2\ell+1}{6\sqrt{3}}=0.866025$&$-\frac{1}{6\sqrt{3}}=-0.0962250$&1 \\
\hline
Massless scalar&$0.867416$&$-0.0963917$&$1.0001$ \\
\hline
Electromagnetic&$0.853095$&$-0.0958599$&$1.0113$ \\
\hline
Metric perturbation&$0.809178$&$-0.0941640$&$1.0473$ \\
\hline
\end{tabularx}
\end{table}

The purpose of this subsection is to compare the results in the time-domain analyses with the frequency domain analyses listed in Table~\ref{tab:frequencies}. 
Therefore, we need to clarify the relation between the master variables $\ee^{-\ii\tilde \Omega T}\psi^{(s)}(x)$ used in the frequency domain analyses with the master variables $\Psi^{(s)}$ defined in this subsection. 
In particular, we focus on the characteristic frequencies listed in Table~\ref{tab:frequencies}, that is, the period and damping rate of the oscillation measured at the radius $x={\rm const}$. 
It should be noted that $x={\rm const.}$ does not imply the constant areal radius $r=2M(\mathcal V)x$, and the areal radius 
increases with time proportionally to $M(\mathcal V)$. 
In the time-domain analyses, we may define the oscillation frequency and the damping rate by monitoring the amplitudes of the master variable at each oscillation peak as follows. 
Let us label each peak of the waveform by an integer $n$, and 
let $\Psi_n$ and $\mathcal V_n$ denote the amplitude and the time at the $n$-th peak.  
Then we can extract the real frequency $\omega^{\rm Re}_n$, imaginary frequency  $\omega^{\rm Im}_n$ and the ratio $\mathcal R_n$ from the following equations:
\begin{eqnarray}
  \omega_n^{\rm Re}&=&\frac{\pi}{\mathcal V_n-\mathcal V_{n-1}}, \\
  \omega_n^{\rm Im}&=&\frac{\ln\left|\Psi^{(s)}_n/\Psi^{(s)}_{n-1}\right|}{\mathcal V_n-\mathcal V_{n-1}}, \\
  \mathcal R_n&:=&\left|\frac{\omega^{\rm Im}_n}{\omega^{\rm Re}_n}\right|=\frac{\ln\left|\Psi^{(s)}_{n-1}/\Psi^{(s)}_n\right|}{\pi}. 
\end{eqnarray}
That is, we have assumed that, in the domain $\mathcal V_{n-1}<\mathcal V\leq\mathcal V_n$, the waveform can be expressed as 
\begin{equation}
    \Psi\sim {\rm Re}[e^{-\ii S({\mathcal  V})}]
\end{equation}
with
\begin{equation}
    S=S(\mathcal V_{n-1})+(\omega_n^{\rm Re}+\ii \omega_n^{\rm Im})(\mathcal V-\mathcal V_{n-1})+\mathcal O\left((\mathcal V-\mathcal V_{n-1})^2\right), 
\end{equation}
where the higher order terms of $\mathcal O\left((\mathcal V-\mathcal V_{n-1})^2\right)$ are
assumed to be negligible as in the standard WKB approximation. 
Under this approximation, the complex frequency is consistent with the definition 
\begin{equation}
    \omega^{\rm Re} + \ii \omega^{\rm Im} =dS/d{\mathcal  V}. 
\end{equation}

\begin{figure}[h!]
\centering
\includegraphics[scale=1.]{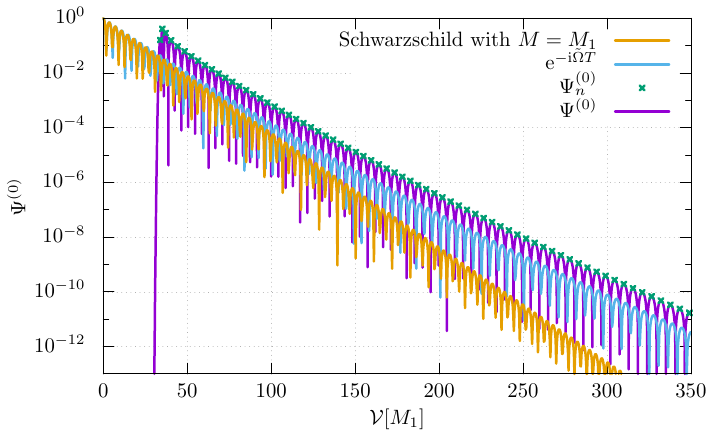}
\caption{\baselineskip5mm
The waveform measured at $x=60$ for the massless scalar field ($s=0$) and $\ell=4$ with $M'=\pi/2000$. 
For the Schwarzschild case, we plot the waveform $\propto \ee^{-\ii\omega_{\rm Sch}\mathcal V}$ with the real and imaginary frequency values listed in Tabel~\ref{tab:PLSch} for $M=M_1$.
The waveforms $\propto \ee^{-\ii\omega_{\rm Sch} \mathcal V}$ and $\propto \ee^{-\ii\hat\Omega T(\mathcal V)}$ are normalized so that they will become unity at $\mathcal V=0$. 
}
\label{fig:damp_linearVaidya}
\end{figure}

We solve the master equations together with Eq.~\eqref{eq:rUV} in the time domain. 
The initial condition is set on a $\mathcal V=\mathcal V_{\rm ini}$ null surface as a function of $r(\mathcal U,\mathcal V_{\rm ini})$ as 
\begin{equation}
  \Psi=\mathcal A\exp\left[-\frac{\left(r(\mathcal U,\mathcal V_{\rm ini})-3M_1\right)^2}{2\sigma_r^2}\right], 
\end{equation}
where $\mathcal A$ and $\sigma_r$ are set to $0.7$ and $M_1/10$. 
The waveform measured at $x=60$ for the massless scalar field ($s=0$) and $\ell=4$ 
with $M'=\pi/2000$ 
is shown 
in Fig.~\ref{fig:damp_linearVaidya}. 
  In Fig.~\ref{fig:damp_linearVaidya}, for the Schwarzschild case, we plot the waveform $\propto \ee^{-\ii\omega_{\rm Sch}\mathcal V}$ with the real and imaginary frequency values listed in Table~\ref{tab:PLSch} for $M=M_1$.
The waveforms $\propto \ee^{-\ii\omega_{\rm Sch} \mathcal V}$ and $\propto \ee^{-\ii\hat\Omega T(\mathcal V)}$ are normalized 
in such a way that they will become unity at $\mathcal V=0$. 

For the massless scalar field, focusing on the time-dependence with a fixed value of $x$, we find 
\begin{equation}
  \Psi^{(0)}\sim r\Phi\sim x\tilde \Phi \sim \ee^{-\ii\tilde \Omega T}\psi^{(0)}(x)= \ee^{-\ii\hat \Omega T}u(x). 
\end{equation}
For the Maxwell field, we find 
\begin{equation}
  \Psi^{(1)}= \ee^{-\ii\tilde \Omega T}\psi^{(1)}(x)= \ee^{-\ii\hat \Omega T}\psi^{(1)}(x). 
\end{equation}
We need a little more consideration for the metric perturbation. 
First, from the coordinate relations, we find 
\begin{eqnarray}
  \dd x&=&\frac{\tilde f(x)}{4M}\dd \mathcal V-\frac{\ee^{\sigma}}{4M}\dd \mathcal U, 
  \label{eq:dxdvdu}\\
\dd T&=&\frac{1}{4M}\dd \mathcal V+\frac{\ee^\sigma}{4M\tilde f(x)}\dd \mathcal U.
\label{eq:dTdvdu} 
\end{eqnarray} 
Comparing the expressions for the metric perturbations \eqref{eq:tilh} and \eqref{eq:hH}, we find
\begin{equation}
  4M^2\ee^{-\ii\tilde \Omega T}\left(h_0(x)\dd T+h_1(x)\dd x\right)=H_0(\mathcal U,\mathcal V)\dd \mathcal U+H_1(\mathcal U,\mathcal V) \dd \mathcal V. 
\end{equation}
Then, by using the relations \eqref{eq:dxdvdu} and \eqref{eq:dTdvdu}, we find 
\begin{eqnarray}
  \ee^{-\ii\tilde \Omega T}h_1=(2M)^{-1}\left(H_1-\frac{\tilde f(x)}{\ee^{\sigma}}H_0\right). 
\end{eqnarray}
Substituting the expressions \eqref{eq:H0} and \eqref{eq:H1} and using the transformation of the partial derivatives
\begin{eqnarray}
  \ee^{-\sigma}\frac{\del\Psi^{(2)}}{\del \mathcal U}&=&\frac{1}{4M\tilde f}\left(\frac{\del\Psi^{(2)}}{\del T}-\tilde f\frac{\del\Psi^{(2)}}{\del x}\right),\\
  \frac{\del\Psi^{(2)}}{\del \mathcal V}&=&\frac{1}{4M}\left(\frac{\del\Psi^{(2)}}{\del T}+\tilde f\frac{\del\Psi^{(2)}}{\del x}\right),  
\end{eqnarray}
we find
\begin{equation}
  \ee^{-\ii\tilde \Omega T}\psi^{(2)}=(4M)^{-1}\left(1-\frac{1}{x}-2M'x+x\tilde f(x)\frac{\del}{\del x}\right)\Psi^{(2)}\sim \left(1+M'\frac{\mathcal V}{M_1}\right)^{-1}\Psi^{(2)}, 
\end{equation}
where we have evaluated the time-dependent part only in the last expression. 
Then, we have
\begin{equation}
  \Psi^{(2)}\sim\ee^{-\ii\tilde \Omega T}\left(1+M'\frac{\mathcal V}{M_1}\right)=\ee^{-\ii\hat \Omega T}. 
\end{equation}

In summary, the time-dependence of the master variables can be rewritten as a function of $\mathcal V$ as 
\begin{equation}
\Psi^{(s)}\sim  \exp\left[-\ii \hat \Omega T\right]
 =\left(1+M'\frac{\mathcal V}{M_1}\right)^{{\rm Im} \hat \Omega/(2M')}\left(1+M'\frac{\mathcal V}{M_1}\right)^{-\ii {\rm Re} \hat\Omega/(2M')}. 
\end{equation}
Under the WKB approximation, the complex frequency can be estimated as 
\begin{equation}
    \omega\sim \left.\frac{\del}{\del \mathcal V}\left(\hat \Omega T\right)\right|_{x={\rm const.}}=\hat \Omega\left.\frac{\del T}{\del \mathcal V}\right|_{x={\rm const.}}= \frac{\hat \Omega}{2M(\mathcal V)}. 
    \label{eq:omOm}
\end{equation}
According to our time-domain analysis, more precisely, the complex frequency can be calculated as follows. 
The oscillation peak can be identified with the condition 
\begin{equation}
  \left.\frac{\del }{\del \mathcal V}{\rm Re}\left(\ee^{-\ii \hat \Omega T}\right)\right|_{\mathcal V=\mathcal V_n}=0, 
\end{equation}
so that
\begin{equation}
  \frac{{\rm Re}\hat \Omega}{2M'}\ln\left|1+M'\frac{\mathcal V_n}{M_1}\right|=\pi n+\arg\hat \Omega.  
\end{equation}
Then, we find 
\begin{eqnarray}
  \omega^{\rm Re}_n&=&\frac{\pi M'}{M_1} \left(1-\exp\left[-\frac{2\pi M'}{{\rm Re}\hat \Omega}{\rm arg}\hat \Omega\right]\right)^{-1}\exp\left[-\frac{2M'}{{\rm Re}\hat \Omega}{\rm arg}\hat \Omega\right]\exp\left[-\frac{2n\pi M'}{{\rm Re}\hat \Omega}\right],\\
  \mathcal R_n&=&\left|\frac{\omega^{\rm Im}_n}{\omega^{\rm Re}_n}\right|=\frac{\ln \left|\Psi^{(s)}_{n-1}/\Psi^{(s)}_n\right|}{\pi}=-\frac{{\rm Im}\hat\Omega}{{\rm Re}\hat\Omega}.   
\end{eqnarray}

\begin{figure}[h!]
\centering
\includegraphics[scale=0.63]{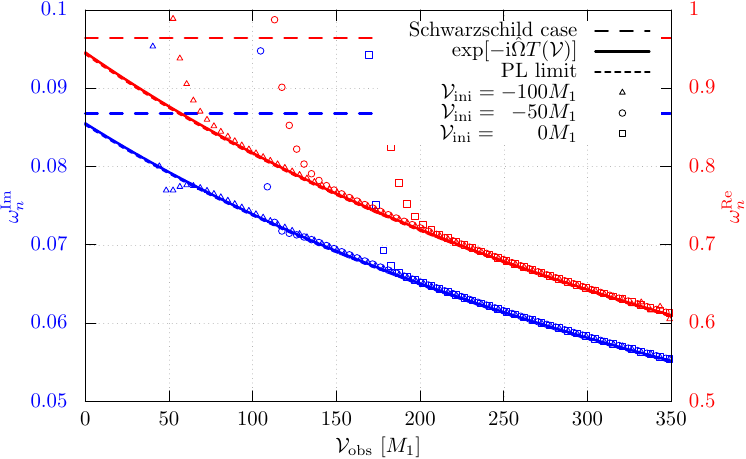}~~
\includegraphics[scale=0.63]{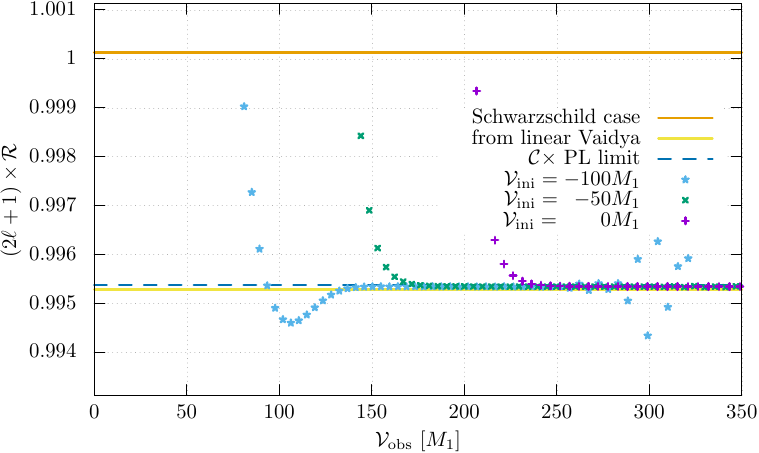}
\includegraphics[scale=0.63]{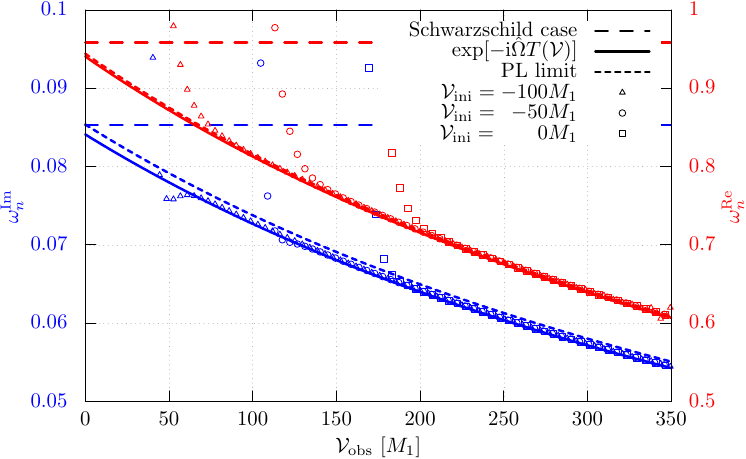}~~
\includegraphics[scale=0.63]{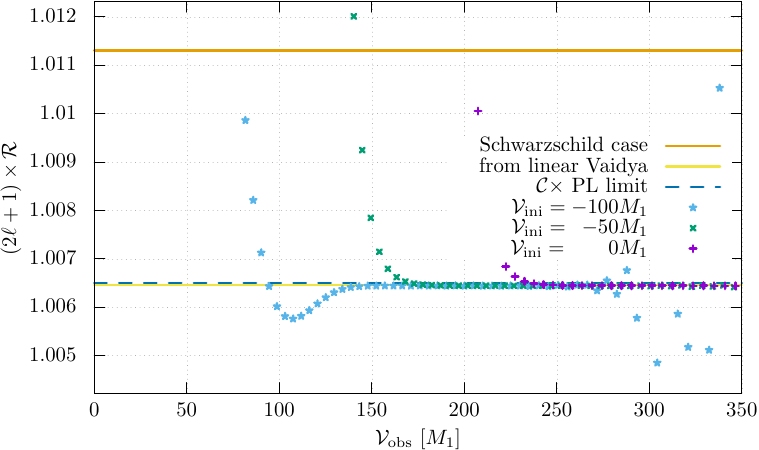}
\includegraphics[scale=0.63]{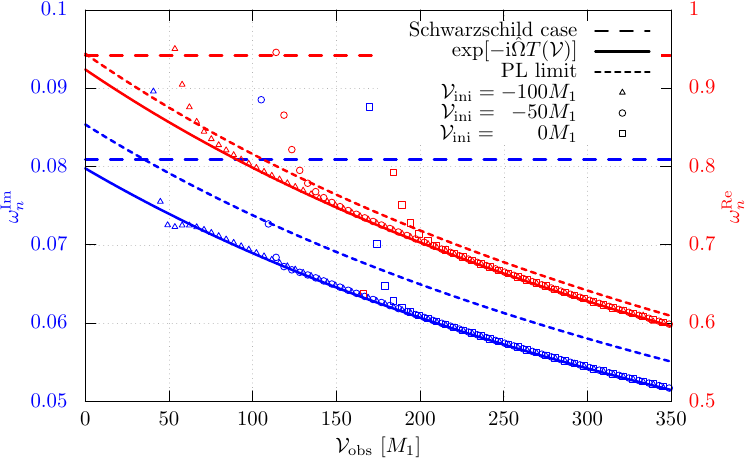}~~
\includegraphics[scale=0.63]{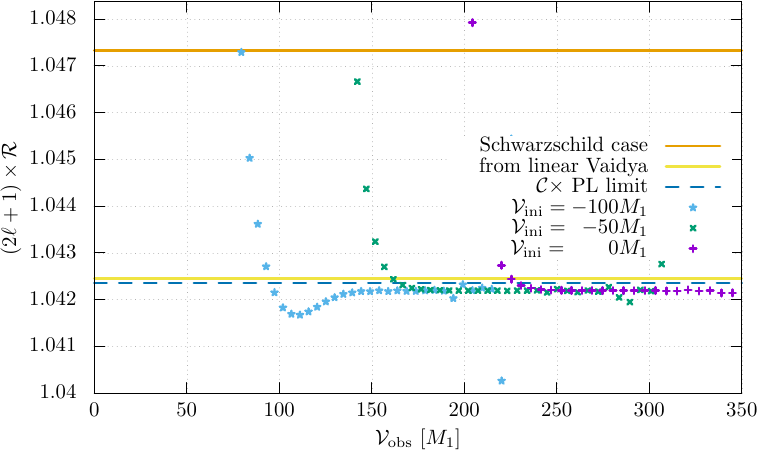}
\caption{\baselineskip5mm
The values of $\omega_n^{\rm Re/Im}$ (left) and $\mathcal R_n$ (right)
for the massless scalar field (top), Maxwell field (middle), and metric perturbation (bottom) with $M'=\pi/2000$ and $\ell=4$. 
In the left panels, the solid lines and short dashed lines show the values of 
$\hat \Omega^{\rm Re/Im}/(2M(\mathcal V))$ and $\Omega^{\rm Re/Im}_{\rm PL}/(2M(\mathcal V))$, respectively (see Eq.~\eqref{eq:omOm} for the relation between $\omega$ and $\hat \Omega$). 
The horizontal dashed line shows the value for the Schwarzschild case with $M=M_1$. 
}
\label{fig:Om_R}
\end{figure}

In Fig.~\ref{fig:Om_R}, we show the values of $\omega_n^{\rm Re/Im}$ and $\mathcal R_n$ 
for the massless scalar field (top), Maxwell field (middle), and metric perturbation (bottom) with $M'=\pi/2000$ and $\ell=4$. 
The amplitude of the damping oscillation drops quickly, so that the numerical accuracy cannot be maintained. 
Therefore, we plot the results with several different values of the initial time $\mathcal V_{\rm ini}$, 
and focus on the time-domain in which the waveform exhibits an idealized exponentially damping oscillation. 
We can find that, in the time-domain analyses, calculating the value of $\mathcal R_n$ along the 
world line of constant $x$, we obtain similar values to those of the frequency domain analyses and the PL analyses with the correction factor $\mathcal C$ estimated from the Schwarzschild case (see Fig.~\ref{fig:Om_R}). 

\section{Penrose limit and ringdown analyses in Vaidya spacetimes with time-dependent accretion rates}
\label{sec:timedepVaidya}
\subsection{Penrose limit}
In this section, we treat the Vaidya spacetime with time-dependent accretion rates. 
To discuss the PL, first, we need to obtain the trajectory of the dynamical photon sphere reported in Ref.~\cite{Koga:2022dsu}. 
For the mass function $M(\mathcal V)$, following Ref.~\cite{Koga:2022dsu}, we consider the following functional form:
\begin{equation}
  M(\mathcal V)=
  \left\{
    \begin{array}{cc}
      M_1&\mathcal V\leq \mathcal V_1\\
      M_1+\frac{1}{2}(M_2-M_1)\left(1-\cos\left(\frac{\mathcal V-\mathcal V_1}{\mathcal V_2-\mathcal V_1}\pi\right)\right)&\mathcal V_1<\mathcal V\leq \mathcal V_2\\
      M_2&\mathcal V_2<\mathcal V
    \end{array}
  \right. .
\end{equation}
We note that the maximum accretion rate $M'(\mathcal V)$ is given by 
\begin{equation}
M'_{\rm max}= M'((\mathcal V_2+\mathcal V_1)/2)=\frac{\pi}{2}\frac{M_2-M_1}{\mathcal V_2-\mathcal V_1}. 
\end{equation}

\begin{figure}[h!]
  \centering
  \includegraphics[scale=1.2]{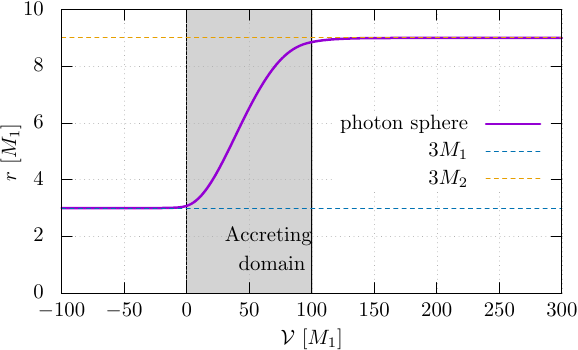}
  \caption{
    \baselineskip5mm
    The photon sphere trajectory for the Vaidya spacetime with 
  $M_2=3M_1$, $\mathcal V_1=0$ and $\mathcal V_2=100M_1$. The null dust is accreting in the shaded region. 
The value of the maximum accretion rate $M'_{\rm max}$ is given by $\pi/100$. 
  }
  \label{fig:trajectory}
  \end{figure}

For the metric \eqref{eq:metVaidya}, null geodesic equations on the equatorial plane $\theta=\pi/2$ 
can be derived from the following Lagrangian:  
\begin{equation}
  \mathcal L=\frac{1}{2}g_{\mu\nu}\frac{\dd x^\mu}{\dd\tau}\frac{\dd x^\nu}{\dd\tau}=\frac{1}{2}\left[-\left(1-\frac{2M(\mathcal V)}{r}\right)\left(\frac{\dd \mathcal V}{\dd \tau}\right)^2+2\frac{\dd \mathcal V}{\dd \tau}\frac{\dd r}{\dd \tau}+r^2\left(\frac{\dd \phi}{\dd \tau}\right)^2\right]. 
\end{equation}
The generalized momenta are given by 
\begin{eqnarray}
  \frac{\del \mathcal L}{\del \dot r}&=&\frac{\dd \mathcal V}{\dd \tau}, \\
  \frac{\del \mathcal L}{\del \dot {\mathcal V}}&=&-\left(1-\frac{2M(\mathcal V)}{r}\right)\frac{\dd\mathcal V}{\dd \tau}+\frac{\dd r}{\dd \tau}=:E_{\mathcal V}, \\
  \frac{\del \mathcal L}{\del \dot \phi}&=&r^2\frac{\dd \phi}{\dd \tau}=:L, 
\end{eqnarray}
where $L$ is a constant of motion while $E_{\mathcal V}$ is not a constant. 
The equations of motion and constraint are written as follows:
\begin{eqnarray}
  \frac{\dd ^2 \mathcal V}{\dd \tau^2}&=&-\frac{M(\mathcal V)}{r^2}\left(\frac{\dd \mathcal V}{\dd \tau}\right)^2+\frac{L^2}{r^3},\\
  \frac{\dd^2 r}{\dd \tau^2}&=&\left(1-\frac{2M(\mathcal V)}{r}\right)\frac{\dd^2 \mathcal V}{\dd \tau^2}-\frac{M'(\mathcal V)}{r}\left(\frac{\dd \mathcal V}{\dd \tau}\right)^2+\frac{2M(\mathcal V)}{r^2}\frac{\dd \mathcal V}{\dd \tau}\frac{\dd r}{\dd \tau}, \\
  0&=&\left(1-\frac{2M(\mathcal V)}{r}\right)\left(\frac{\dd \mathcal V}{\dd \tau}\right)^2-2\frac{\dd\mathcal V}{\dd \tau}\frac{\dd r}{\dd \tau}-\frac{L^2}{r^2}, 
\end{eqnarray}
where $M'(\mathcal V)=\dd M(\mathcal V)/\dd \mathcal V$. 
We can reproduce the dynamical photon sphere by following the shooting method described in Ref.~\cite{Koga:2022dsu} for $M_2=3M_1$, $\mathcal V_1=0$, and $\mathcal V_2=100M_1$ (Fig.~\ref{fig:trajectory}). 
In this case, the value of $M'_{\rm max}$ is given by $\pi/100$. 
We plot the trajectory of the photon sphere given by a function of the time coordinate $\mathcal V$ as $r_{\rm PS}(\mathcal V)$ in 
Fig.~\ref{fig:trajectory}. 

\begin{figure}[h!]
  \centering
  \includegraphics[scale=1.]{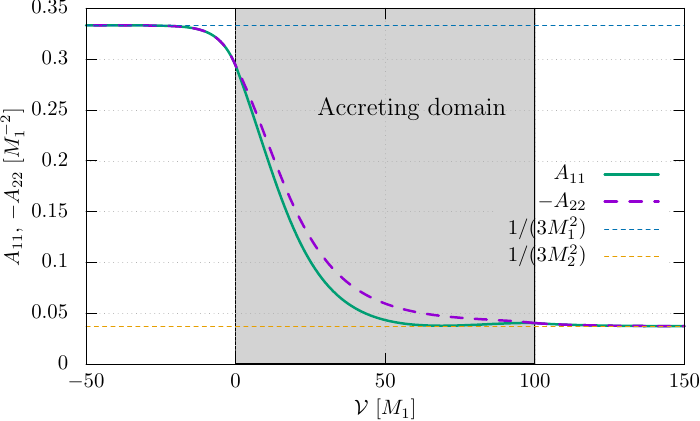}
  \caption{
    \baselineskip5mm
    The values of $A_{11}$ and $-A_{22}$ with 
  $M_2=3M_1$, $\mathcal V_1=0$ and $\mathcal V_2=100M_1$. The null dust is accreting in the shaded region. 
  }
  \label{fig:A11A22}
  \end{figure}

  Along a null geodesic lying on the photon sphere, we calculate parallelly transported zweibeins, 
  satisfying 
  \begin{equation}
    k^\mu\nabla_\mu E^\mu_{(a)}=0. 
  \end{equation}
  Then the functions $A_{ab}$ contained in the metric components shown in Eq.~\eqref{eq:metBrink} 
can be calculated by using Eq.~\eqref{eq:AabfromE}. 
We plot $A_{11}$ and $-A_{22}$ in Fig.~\ref{fig:A11A22} for $M_2=3M_1$, $\mathcal V_1=0$ and $\mathcal V_2=100M_1$.
%
  We may consider the adiabatic approximation for this dynamical spacetime of the PL. 
  Specifically, in the small domain around each time, the geometry can be approximated by 
  the metric \eqref{eq:PLlimit_sp}. 
  Then, we can effectively define $\lambda$ and $\Omega$ from the correspondence \eqref{eq:corresp} as follows:
  \begin{eqnarray}
    \lambda_{\rm ad}:=\frac{\alpha_{\rm ad} E_{\mathcal  V}}{\beta_{\rm ad} L}, \\
    \Omega_{\rm ad}:=\frac{E_{\mathcal  V}}{L},  
  \end{eqnarray}
  where 
  \begin{eqnarray}
    \alpha_{\rm ad}&=&\sqrt{A_{11}}, \\
\beta_{\rm ad}&=&\sqrt{-A_{22}}. 
  \end{eqnarray}
  Here it should be noted that $E_{\mathcal V}$ is not a constant of motion, and 
  we normalize it by imposing $E_{\mathcal V}=1$ at a sufficiently early time. 
  The effective values of the frequencies and those ratios can be defined by 
  \begin{eqnarray}
    \omega^{\rm Re}_{\rm PL,ad}&:=&\frac{\Omega_{\rm ad}}{2}\left(2\ell+1\right), \\
    \omega^{\rm Im}_{\rm PL,ad}&:=&\frac{\lambda_{\rm ad}}{2}, \\
    \mathcal R^{\rm ad}_{\rm PL}&:=&\left|\frac{\omega^{\rm Im}_{\rm PL,ad}}{\omega^{\rm Re}_{\rm PL,ad}}\right|. 
  \end{eqnarray}

  These quantities are defined on the photon sphere trajectory, and can be regarded as functions of $\mathcal V$. 
  We can also define the time of observation $\mathcal V_{\rm obs}$ 
  by solving radial geodesics from points on the photon sphere to 
  a given radius $r_{\rm obs}$. 
  The observer at $r=r_{\rm obs}$ can, in principle, detect the effects of the photon sphere dynamics from $\mathcal V=\mathcal V_{\rm obs}$. 
  That is, we define $\mathcal V_{\rm obs}$ as a function of $\mathcal V$ such that the radial outward null geodesic 
  emanated at the event $(\mathcal V, r=r_{\rm PS}(\mathcal V))$ reaches 
  the observer radius $r_{\rm obs}$ at $\mathcal V=\mathcal V_{\rm obs}$.

\begin{figure}[h!]
  \centering
  \includegraphics[scale=0.7]{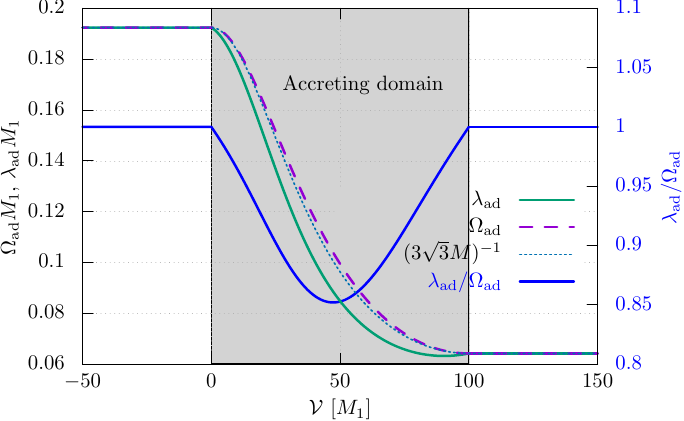}
  \includegraphics[scale=0.7]{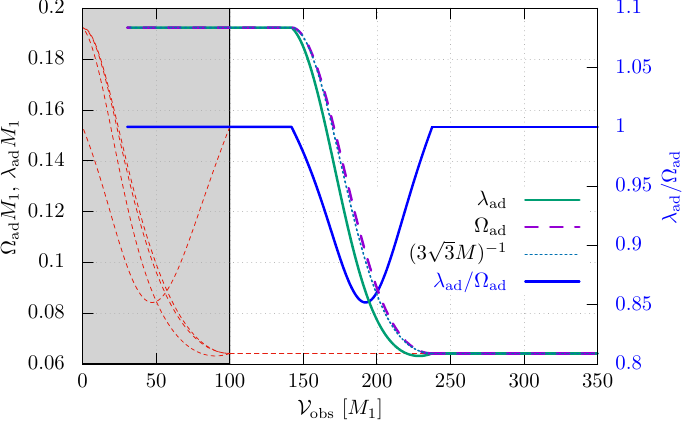}
  \caption{
    \baselineskip5mm
    The values of $\lambda_{\rm ad}$ and $\Omega_{\rm ad}$ with 
  $M_2=3M_1$, $\mathcal V_1=0$ and $\mathcal V_2=100M_1$ as a function of $\mathcal V$ (left panel) and $\mathcal V_{\rm obs}$ (right panel). The null dust is accreting in the shaded region. 
  The ratio $\lambda_{\rm ad}/\Omega_{\rm ad}$ is also plotted with the scale of the right-hand vertical axis as a reference. 
  The red dashed lines in the right panel show all lines in the left panel as references. 
  }
  \label{fig:PP}
  \end{figure}

  The values of $\lambda_{\rm ad}$ and $\Omega_{\rm ad}$ are plotted in Fig.~\ref{fig:PP} as a function of $\mathcal V$ (left panel) and $\mathcal V_{\rm obs}$ with $r_{\rm obs}=60M_1$ (right panel). 
We can find that the ratio $\lambda_{\rm ad}/\Omega_{\rm ad}$ is smaller than 1 
in the accreting domain. 
Therefore, if the feature is reflected in the ringdown waveform, we may observe a temporary reduction in the value of 
$\mathcal R_n(2\ell+1)$ from the value of the Schwarzschild case. 
It would be interesting to note that we trivially obtain the Schwarzschild metric, taking the adiabatic approximation for the original Vaidya metric, since we ignore the time-dependence of the mass function. 
  Then we may expect the constant value of the ratio between $\Omega$ and $\lambda$. 
However, in the adiabatic limit of the PL geometry, the ratio is not constant. 
That is, the adiabatic limit of the PL geometry reflects more details of the dynamical geometry 
compared to the naive adiabatic limit of the original geometry. 

\subsection{Ringdown in Vaidya spacetimes and comparison with the Penrose limit analysis}

\begin{figure}[h!]
  \centering
  \includegraphics[scale=1.]{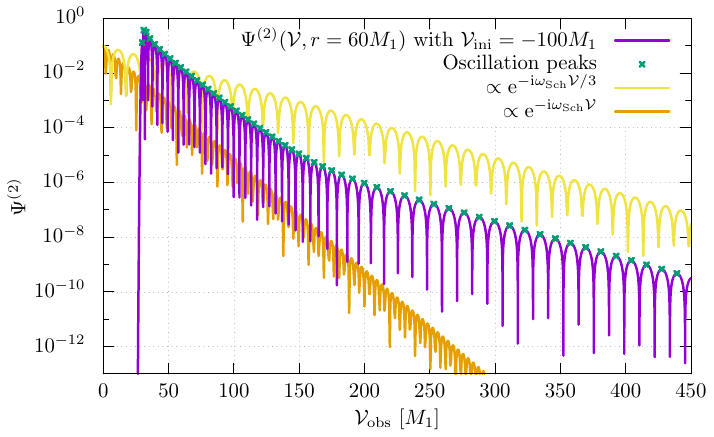}
  \caption{
    \baselineskip5mm
    The waveform of the damping oscillation observed at $r=60M_1$ for $M_2=3M_1$, $\mathcal V_1=0$, $\mathcal V_2=100M_1$, $\mathcal V_{\rm ini}=-100M_1$ and $\ell=4$.  
    }
  \label{fig:damp}
  \end{figure}

First, let us make a demonstration focusing on the case $M_2=3M_1$, in which the effect of the accretion is noticeable. 
We show the waveform $\Psi^{(2)}$ of the damping oscillation observed at $r=60M_1$ for $M_2=3M_1$, $\mathcal V_1=0$, $\mathcal V_2=100M_1$, $\mathcal V_{\rm ini}=-100M_1$ and $\ell=4$ in Fig.~\ref{fig:damp}.  

\begin{figure}[h!]
  \centering
    \includegraphics[scale=0.63]{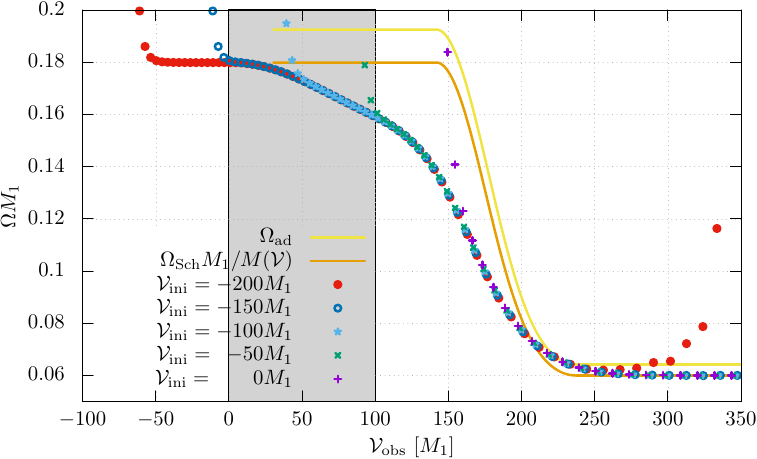}
    \includegraphics[scale=0.63]{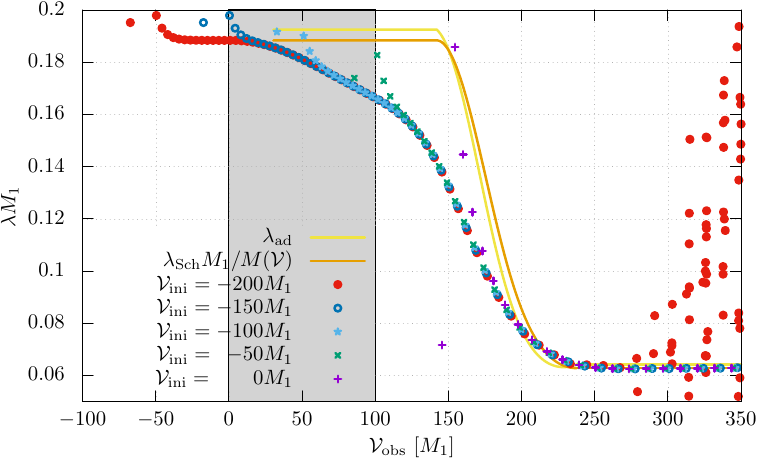}
    \includegraphics[scale=0.8]{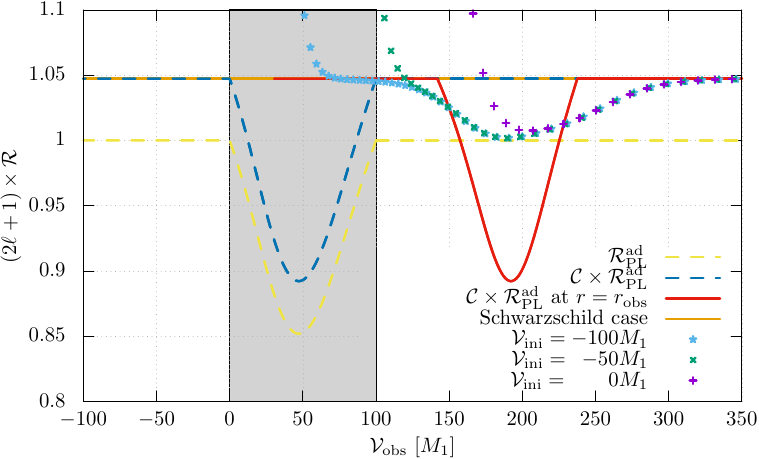}
    \caption{
    \baselineskip5mm
    The values of $\mathcal R_n$, $\Omega_n$ and $\lambda_n$ extracted from the waveform of the damping oscillation observed at $r=60M_1$ for $M_2=3M_1$, $\mathcal V_1=0$, $\mathcal V_2=100M_1$ and $\ell=4$. 
      In each panel, different kinds of dots show the results obtained from the numerical simulation for different values of $\mathcal V_{\rm ini}$ indicated in the legend. 
In the top panels, the yellow lines show the values of $\Omega_{\rm ad}$ and $\lambda_{\rm ad}$, respectively, obtained by the PL analysis. The orange solid lines show the values of 
    $\Omega_{\rm Sch}M_1/M(\mathcal V)$ and $\lambda_{\rm Sch}M_1/M(\mathcal V)$ with $\Omega_{\rm Sch}=2\omega_{\rm Sch}^{\rm Re}/(2\ell+1)$ and $\lambda_{\rm Sch}=2\omega_{\rm Sch}^{\rm Im}$. 
    In the bottom panel, 
      the yellow dashed and orange solid lines show the predictions of the PL analysis $(2\ell+1)\mathcal R_{\rm PL}$ and 
    corresponding Schwarzschild case, respectively. 
    For the bottom panel, since the predicted value from the PL analysis does not match the Schwarzschild case in the early and late times, we multiply the correction factor (blue dashed line) so that it can overlap with the Schwarzschild case in the early and late times. 
    The solid red line shows the corrected PL prediction with the time shifted to that at which the radial null geodesic 
    emanated from the corresponding point on the photon sphere reaches $r=r_{\rm obs}=60M_1$.     
    }
  \label{fig:ratio}
  \end{figure}

We plot the values of $\Omega_n:=2\omega^{\rm Re}_n/(2\ell+1)$, $\lambda_n:=2\omega^{\rm Im}_n$ and $\mathcal R_n$,   
with $\mathcal V_{\rm obs}(\mathcal V_n)$ being the horizontal axis value in Fig.~\ref{fig:ratio}.  
  As in the case of Fig.~\ref{fig:Om_R}, the amplitude of the damping oscillation drops quickly, so that the numerical plots in the late time scatter a lot (see red dots in Fig.~\ref{fig:ratio}) due to the numerical error. 
Therefore, we plot the results with several different values of the initial time $\mathcal V_{\rm ini}$, 
and focus on the time-domain in which the waveform exhibits an idealized exponentially damping oscillation. 
In the top 
panels, we can find that the values of $\Omega_n$ and $\lambda_n$ significantly deviate from $\Omega_{\rm ad}$ and $\lambda_{\rm ad}$, respectively. 
Since the qualitative behaviors are similar in both $\Omega_n$ and $\lambda_n$, these deviations mainly originate from the redshift effect, which equally causes the time delay to the real and imaginary parts of the frequency during the propagation from the UCOP to the observer. 
To extract the effects independently of the redshift, let us focus on the ratio $\mathcal R_n$ between the imaginary and real frequencies shown in the bottom panel of Fig.~\ref{fig:ratio}. 
For the bottom panel, 
the predicted value from the PL analysis does not match the Schwarzschild case in both of the early and late times because of the deviation presented in Table~\ref{tab:PLSch}. 
Thus, 
we multiply the correction factor $\mathcal C$ (blue dashed line), so that it overlaps the Schwarzschild case in the early and late times. 
In the bottom panel of Fig.~\ref{fig:ratio}, we can find that the sequence of $\mathcal R_n$ also has a 
qualitatively similar behavior to the PL prediction $\mathcal R_{\rm PL}$. 
However, the significance of the deviation from the constant value is quantitatively about 3 times different from $\mathcal R_{\rm PL}$. 
In the remaining part of this paper, we carefully investigate the parameter dependence of $\mathcal R_n$, and compare it to $\mathcal R_{\rm PL}$. 

\begin{figure}[h!]
\begin{center}
\includegraphics[scale=0.9]{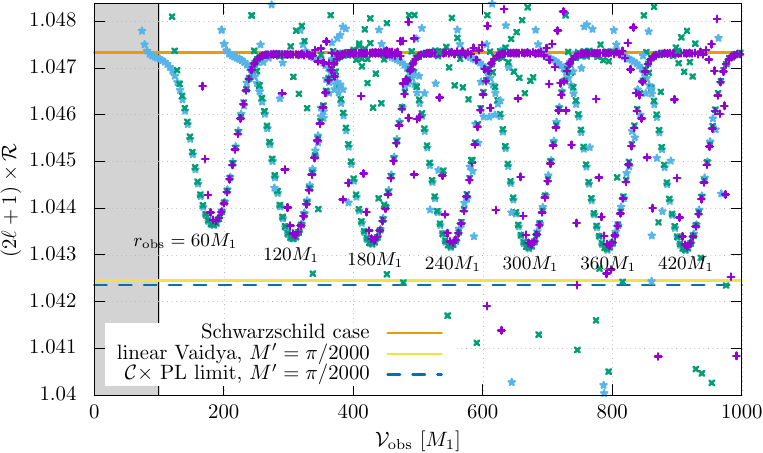}
\caption{
      \baselineskip5mm
$r_{\rm obs}$ dependence of the values of $\mathcal R_n$. 
The values of $\mathcal R_n$ are plotted for 
  $\mathcal V_{\rm ini}=-100M_1$ (blue), $-50M_1$ (green)  and $0$ (purple)} for each value of $r_{\rm obs}$. The orange line shows the value in the Schwarzschild case. 
The yellow line shows the corresponding value 
obtained from the numerical calculation for the static conformal metric of 
the Vaidya spacetime with the constant accretion rate $M'=\pi/2000$. 
The dashed blue line shows the value given by the PL analysis with the constant accretion rate $M'=\pi/2000$. 
\label{fig:robs_conv}
\end{center}
\end{figure}

\begin{figure}[h!]
\begin{center}
\includegraphics[scale=0.63]{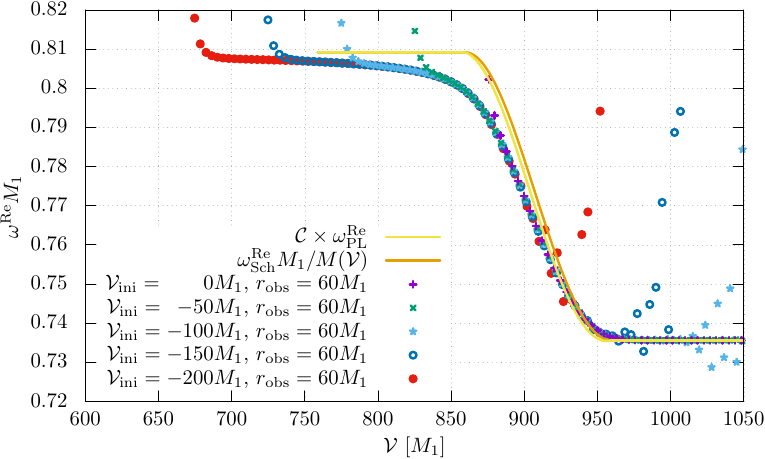}
\includegraphics[scale=0.63]{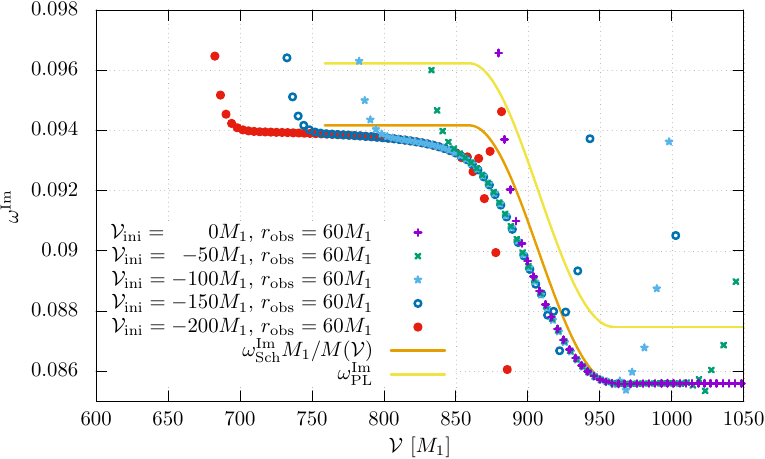}
\caption{    \baselineskip5mm 
The real and imaginary frequencies for $M_2=1.1M_1$. 
}
\label{fig:omega_M211}
\end{center}
\end{figure}

We calculate the values $\Omega_{\rm ad}$ and $\lambda_{\rm ad}$, and 
connect them to an effective complex frequency $\omega_{\rm PL,ad}$. 
Therefore, for $\omega_{\rm PL,ad}$ to be a good approximation, 
the adiabatic approximation must be valid. 
Since the value $M_2=3M_1$ in the previous example is likely to be too large, 
let us consider the case $M_2=1.1M_1$, $\mathcal V_1=0$, $\mathcal V_2=100M_1$, and $\ell=4$ as a reference. 
First, let us check the dependence of the results on the value of $r_{\rm obs}$. 
In Fig.~\ref{fig:robs_conv}, we show the values $\mathcal R_n$. 
One can find that the sequential forms 
slightly depend on the value of $r_{\rm obs}$, and almost convergent for $r_{\rm obs}=420M_1$. 
We note that the $r_{\rm obs}$ dependence does not appear for the standard Schwarzschild ringdown regime in the late times. 
However, when the characteristic frequencies are time-dependent, to be exact, the ringdown waveform should be described by a superposition of the waves 
in a finite frequency range. 
In such general cases, the waveform at a finite radius has a correction term proportional to $1/r_{\rm obs}$ (see, e.g., \cite{Lousto:2010qx}). 
This effect can be neglected for a sufficiently large value of $r_{\rm obs}$, and the result converges. 
Hereafter, we set $r_{\rm obs}=420M_1$. 
The real and imaginary frequencies are plotted in Fig.~\ref{fig:omega_M211} for $M_2=1.1M_1$. 

\begin{figure}[h!]
  \centering
    \includegraphics[scale=0.63]{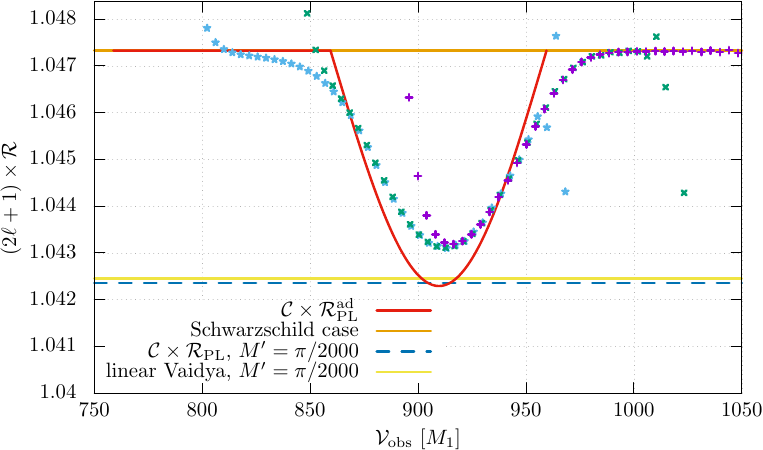}
    \includegraphics[scale=0.63]{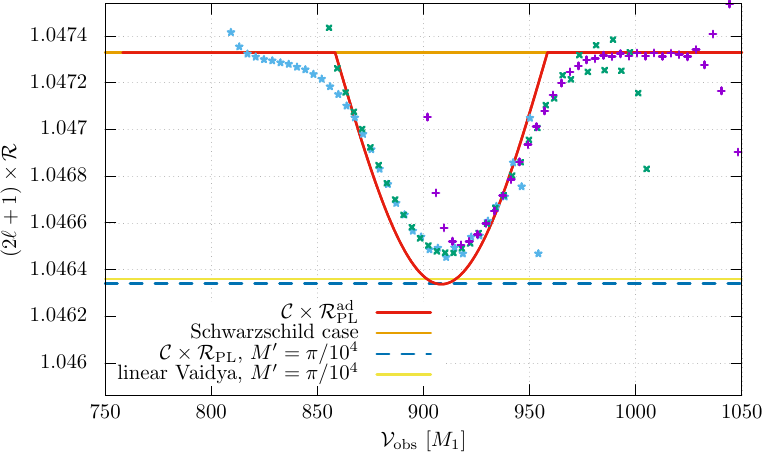}
    \caption{
    \baselineskip5mm
    The values of $\mathcal R_n$ extracted from the waveform of the damping oscillation observed at $r=420M_1$ for $\mathcal V_1=0$, $\mathcal V_2=100M_1$, $\ell=4$. The values of $\mathcal R_n$ are plotted for 
  $\mathcal V_{\rm ini}=-100M_1$ (blue), $-50M_1$ (green)  and $0$ (purple). 
    The left and right panels are for $M_2=1.1M_1$ and $M_2=1.02M_1$, respectively. 
    }
  \label{fig:ratio_M211}
  \end{figure}

\begin{figure}[h!]
\begin{center}
\includegraphics[scale=0.63]{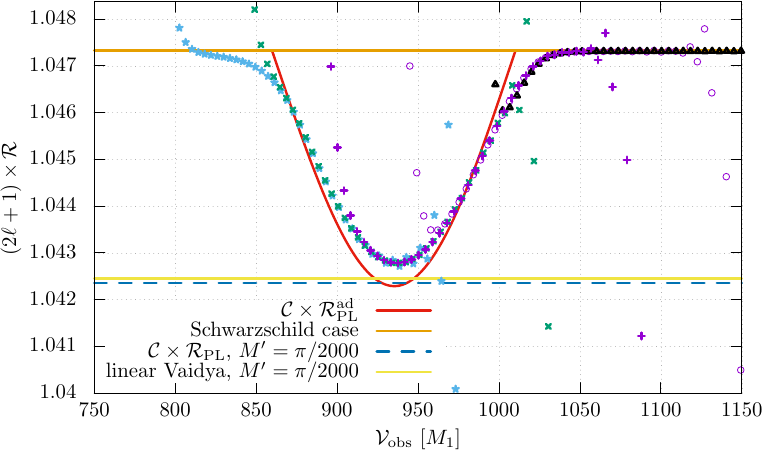}
\includegraphics[scale=0.63]{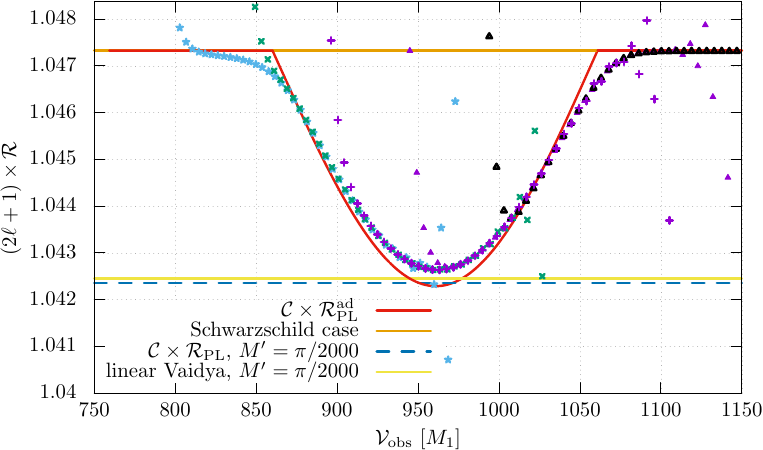}
\caption{\baselineskip5mm 
    The values of $\mathcal R_n$ extracted from the waveform of the damping oscillation observed at $r=420M_1$ for $\mathcal V_1=0$, $M'_{\rm max}=\pi/2000$, $\ell=4$. 
        The values of $\mathcal R_n$ are plotted for 
  $\mathcal V_{\rm ini}=-100M_1$ (blue), $-50M_1$ (green)  and $0$ (purple). 
    The left and right panels are for $\mathcal V_2=150M_1$ and $\mathcal V_2=200M_1$, respectively. 
}
\label{fig:ratio_V2}
\end{center}
\end{figure}

Let us check the dependence of the result on the value of $M_2$. 
The results for the cases $M_2=1.1M_1$ and $M_2=1.02M_1$ are shown in Fig.~\ref{fig:ratio_M211}. 
We find that the feature quantitatively approaches the prediction from the PL analysis 
compared to the case $M_2=3M_1$. 
However, the result does not asymptote to the PL prediction only by changing the value of $M_2$. 
Since the time-domain calculation agrees with the frequency domain and PL analyses with the correction factor $\mathcal C$ in the case of constant accretion rate, this disagreement is expected to be caused by the time-dependent accretion rate during the wave propagation between the UCOP and the observer. 
That is, the difference between the PL limit prediction and the time-domain results is 
mainly caused by the scattering process during the propagation, and cannot be captured by a local geometry around the UCOP. 
Then, let us change the period of the accretion, keeping the value of the maximum accretion rate $M'_{\rm max}$. 
The results for the cases $\mathcal V_2=150M_1$ and $\mathcal V_2=200M_1$ with $M'_{\rm max}=\pi/2000$ are shown in Fig.~\ref{fig:ratio_V2}. 
For a sufficiently large value of $\mathcal V_2$, 
the values of $\mathcal R_n$ extracted from the waveform of the damping oscillation 
seems to approach the prediction from the PL analysis with the correction factor $\mathcal C$. 
  Once we accept the existence of the limit at which the results asymptote to the PL prediction, we may expect that the deviation from the PL prediction would be caused by 
some kind of scattering effect during the wave propagation. 
Then, our results suggest that 
the significance of this scattering effect decreases for a longer period of the accretion with a fixed value of $M'_{\rm max}$. 

\begin{figure}[h!]
  \centering
    \includegraphics[scale=0.63]{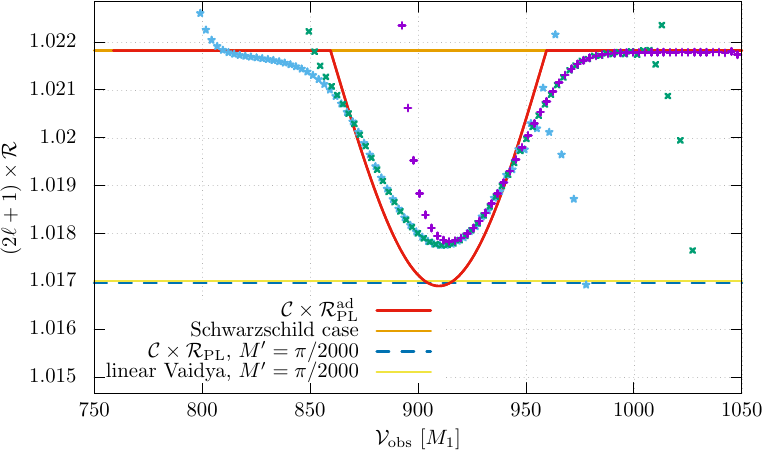}
    \includegraphics[scale=0.63]{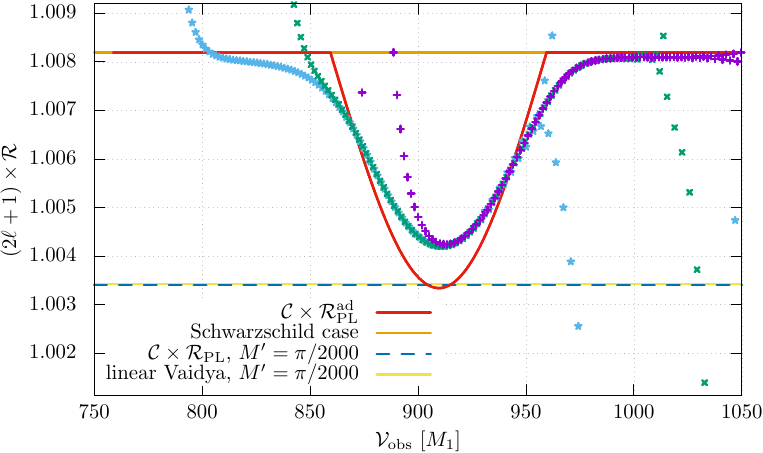}
    \caption{
    \baselineskip5mm
    The values of $\mathcal R_n$ extracted from the waveform of the damping oscillation observed at $r=60M_1$ for $\mathcal V_1=0$, $\mathcal V_2=100M_1$, and $M_2=1.1M_1$. 
        The values of $\mathcal R_n$ are plotted for 
  $\mathcal V_{\rm ini}=-100M_1$ (blue), $-50M_1$ (green)  and $0$ (purple). 
    The left and right panels are for $\ell=6$ and $\ell=10$, respectively. 
    }
  \label{fig:ratio_M211_ell6}
  \end{figure}

Finally, let us check the $\ell$ dependence. 
We might expect a better agreement between the PL prediction and the ringdown waveform or higher values of $\ell$. 
However, for $\ell=6$ and $10$, we obtain similar results to the $\ell=4$ case (see Fig.~\ref{fig:ratio_M211_ell6}). 
Our results indicate that the value $\ell=4$ is sufficiently large 
so that the results will almost converge to the high-frequency limit. 
Thus, in the dynamical case shown here, for the realization of the PL limit, 
the large $\ell$ limit is not a sufficient condition, but we need additional conditions for 
the accretion rate and period. 

\section{Summary and conclusions}
\label{sec:summary_conclusions}

For the 
stationary  
background spacetime, the ringdown waves are known to originate from the damping oscillation modes 
around the UCOP. 
This behavior can be justified considering QNMs in the PL geometry 
around the null geodesic on the UCOP, as is reviewed in Sec.~\ref{sec:QNMPL}. 
The PL geometry just provide us the information of the local geometry near the UCOP. 
Therefore, the scattering effects during the propagation from UCOP to the observer cannot be captured in the PL analyses. 
Nevertheless, in the case of 
stationary 
spacetimes, considering the QNM frequency associated with the 
stationary 
Killing observer, the result can be directly interpreted as the waveform measured by a distant observer. 
Therefore, the PL analyses can be also easily compared with the full QNM analyses. 

Our main purpose was to clarify whether this observation also applies to dynamical spacetimes. 
In this paper, we carefully analyzed the ringdown waveforms in a Vaidya spacetime with time-independent 
(Sec.~\ref{sec:linearVaidy}) and dependent accretion rates (Sec.~\ref{sec:timedepVaidya}), and compared them with the prediction from the PL analyses. 
In the case of a constant accretion rate, the spacetime metric can be described as a conformally static form. 
Then, both the frequency domain QNM analysis and the PL analysis can be performed in the conformal static spacetime. We found that these analyses agree with each other at a certain level when the correction term evaluated by using the Schwarzschild case is taken into account. 
We have also compared the evaluated QNM frequencies with those from the time-domain analyses performed in the full Vaidya spacetime with the observer trajectory adjusted to the frequency domain analysis, and found good agreement again. 
This observation shows us that, even under the existence of the accretion, the PL analysis may give us the local geometrical origin of the ringdown waveforms. 

The cases of time-dependent accretion rate are more complex because of the non-trivial scattering effects during the propagation. 
To extract an adiabatic behavior of the ringdown waveform from the time-domain calculation, 
we have defined the temporal complex frequency by using the neighboring two peaks of the waveform. 
That is, from the time interval and the decay rate between the two peaks, 
we can calculate the real and imaginary parts of the frequency. 
Differently from the constant accretion case, with a finite accretion period, the future asymptotic region is described by the Schwarzschild spacetime of the mass $M_2$. 
However, the ringdown waves emanated from the vicinity of the dynamical photon sphere 
cannot be characterized by the specific complex frequency in the Schwarzschild spacetime with the mass $M_2$
because the radius of the dynamical photon sphere is time-dependent. 
Therefore, we had to carefully check the scattering effects during the propagation. 

The temporal frequency monotonically changes from the Schwarzschild value with the initial mass $M_1$ 
to that with the final mass $M_2$. 
This behavior is shared with the PL analysis, but we observed substantial deviation from the PL prediction in the time dependence of the temporal frequency. 
This substantial deviation is mainly caused by the scattering effects during the propagation in a broad sense 
including the redshift effect. 
The redshift effect is the dominant effect and equally shifts the real and imaginary parts of the frequency. 
Since the redshift effect is an additional effect to the original emission process around the photon sphere, the redshift effect causes a deviation between the PL prediction and 
the actual waveform. 
It should be noted that, in a 
stationary 
case, the redshift effect is trivial and automatically 
taken into account by considering the frequency for the time coordinate associated with the 
stationary 
Killing vector field. 
Therefore, the redshift effect can be also regarded as a peculiar effect in the dynamical spacetime, 
and it contains the information about the accretion. 

Since the redshift equally affects to the real and imaginary parts of the complex frequency, we may extract the purely geometrical information around the dynamical photon sphere by taking 
the ratio between the imaginary ($\omega^{\rm Im}$) and real ($\omega^{\rm Re}$) parts of the frequency. 
We found the characteristic temporal decrease in the ratio $\mathcal R=\omega^{\rm Im}/\omega^{\rm Re}$ both 
in the PL analysis and the time-domain numerical simulation. 
It has been shown that, for sufficiently large angular momentum quantum number $\ell$ ($\ell\geq4$) and observer radius $r_{\rm obs}$ ($r_{\rm obs}\gtrsim 400M_1$), 
the time-dependence of $\mathcal R$ approaches to the PL prediction 
under an additional condition 
$M''M_1\ll M'$ with the prime `` ${}'$ " being the time derivative.  
This result indicates that there is still a remaining scattering effect associated with the time-dependence of the accretion rate, and without this remaining scattering effect in an appropriate limit, the PL prediction may give a good approximation for $\mathcal R$. 
It implies that, in principle, the ringdown waveform can deliver the information of the geometry around the photon sphere to a distant observer even in a dynamical case.

\section*{Acknowledgements}

This work was supported by JSPS KAKENHI Grant Numbers JP25K07281 (C.Y.), JP24K07027 (C.Y.), JP22K03626 (M.K.), JP21H05182 (A.I.), JP21H05186 (A.I.), and JP25K07306 (A.I.). 
R.O. would like to take this opportunity to thank the ``THERS Make New Standards Program for the Next Generation Researchers" supported by JST SPRING, Grant Number JPMJSP2125. 

\appendix

\titleformat{\section}  
  {\fontsize{14}{16}\bfseries} 
  {\Alph{section}.} 
  {0.5em} 
  {} 
  [] 

\titleformat{\subsection}  
  {\fontsize{12}{14}\bfseries} 
  {\arabic{subsection}.} 
  {0.5em} 
  {} 
  [] 



\section{Review of Penrose limit geometry}

\subsection{Penrose limit geometries}
\label{sec:PL}
First, let us briefly review the PL geometries~\cite{Penrose1976} following Ref.~\cite{Blau_review}. 
The procedure to obtain the PL metric around a null geodesic $\gamma$ starts from 
finding the Penrose coordinates adapted to $\gamma$ in which the metric is described by 
\begin{equation}
  \dd s^2_\gamma=2\dd U\dd V+a(U,V,Y^k)\dd V^2+2b_i(U,V,Y^k)\dd Y^i \dd V+g_{ij}(U,V,Y^k)\dd Y^i \dd Y^j, 
  \label{eq:orimet}
\end{equation}
where $i$, $j$ and $k$ run over 1 to 2. 
There is a general way to find this Penrose coordinate based on the Hamilton-Jacobi function $S(x^\mu)$ for null geodesic congruence 
involving $\gamma$, which will be briefly described in the next subsection.  
Readers are asked to refer to Ref.~\cite{Blau_review} for details. 
We will also describe the specific example of the construction of the Penrose coordinates for the unstable circular orbit of photons (UCOP) in a static spherically symmetric spacetime in ~\ref{sec:PLPS}. 
However, since we will take a shortcut to obtain the PL geometry in the main text, the derivation is not necessarily needed for understanding our results in the main text. 

Once the metric in the Penrose coordinates is given, the PL is defined by the following procedure:
\begin{equation}
  \dd \bar s^2=\lim_{\varepsilon\rightarrow0}\left(\varepsilon^{-2}\left.\dd s_\gamma^2\right|_{(U,V,Y^k)\rightarrow(U,\varepsilon^2V,\varepsilon Y^k)}\right). 
  \label{eq:PLlimit}
\end{equation}
The scaling $(U,V,Y^k)\rightarrow(U,\varepsilon^2V,\varepsilon Y^k)$ is the combination of the "boost'' $(U,V,Y^k)\rightarrow(\varepsilon^{-1}U,\varepsilon V,Y^k)$ and the uniform rescale $(U,V,Y^k)\rightarrow \varepsilon(U,V,Y^k)$. 
As a result, we obtain 
\begin{equation}
  \dd \bar s^2=2\dd U\dd V+\bar g_{ij}(U)\dd Y^i \dd Y^j, 
\end{equation}
where $\bar g_{ij}(U)=g_{ij}(U,0,0)=\lim_{\varepsilon\rightarrow0}g_{ij}(U,\varepsilon^2V,\varepsilon Y^k)$. 
This metric form is known as a plane wave metric in Rosen coordinates. 
It is worth noting that the finite region described by the cubic domain of the coordinates $(U, V, Y^k)$ for the PL metric 
corresponds to the infinitely thin region along the geodesic $\gamma$ with a finite length of the affine parameter $U$. 
Therefore, taking the PL, we can extract the geometry in the vicinity of the null geodesic $\gamma$. 

The Rosen coordinate often suffers from coordinate singularities. 
More convenient coordinates are known as Brinkmann coordinates $(u,v,x^a)$ in which the metric is described as 
\begin{equation}
  \dd \bar s^2=2\dd u\dd v+A_{ab}(u)x^ax^b\dd u^2+\delta_{ab}\dd x^a\dd x^b, 
  \label{eq:metBrink}
\end{equation}
where $a$ and $b$ run over 1 to 2 and $\delta_{ab}$ is the Kronecker's delta in the two-dimensional space. 
The coordinate transformation from the Rosen to Brinkmann coordinates is given by 
\begin{eqnarray}
  U&=&u, \\
  V&=&v+\frac{1}{2}\partial_u \bar E^{(a)}_i \bar E^i_{(b)}x^ax^b, \\
  Y^i&=&\bar E^i_{(a)}x^a, 
\end{eqnarray}
where $\bar E^i_{(a)}$ is a zweibein satisfying the symmetry condition: 
\begin{equation}
  \bar E^i_{(b)}\del_u \bar E^{(a)}_i=\bar E^i_{(a)}\del_u \bar E^{(b)}_i. 
\end{equation}
Then the metric component $A_{ab}$ is given by 
\begin{equation}
  A_{ab}=\delta_{ac}\ddot {\bar E}^c_i\bar E_b^j. 
  \label{eq:AabfromE}
\end{equation}
Practically, we can construct the zweibein satisfying the symmetry condition by 
constructing the parallelly transported zweibein in the original metric $ds^2_\gamma$ 
along $\gamma$. 
That is $\bar E^{(a)}_i=E^{(a)}_i$ with 
\begin{eqnarray}
&&g_{\mu\nu}E^i_{(a)}(\del_i)^\nu E^j_{(b)}(\del_j)^\nu=E_j^{(a)}E^j_{(b)}=\delta^a_b,  \\
&&(\del_U)^\mu\nabla_\mu (E^i_{(a)}(\del_i)^\nu)=0. 
\end{eqnarray}
Indeed, we can show that the zweibeins $E^i_{(a)}$ satisfy the symmetry condition 
$E^i_{(b)}\del_U E^{(a)}_i=E^i_{(a)}\del_U E^{(b)}_i$. 

The identification between $\bar E^{(a)}_i$ and $E^{(a)}_i$ enable us to make a covariant description of the PL. 
The covariant description can be made based on the following two facts:
\begin{eqnarray}
  &&\bar R_{UiU}^{~~~~j}=\left.R_{UiU}^{~~~~j}\right|_\gamma, \\
&&A_{ab}=-\bar R_{UiUj}\bar E^i_{(a)}\bar E^j_{(b)}=-\left.\left(R_{UiUj} E^i_{(a)} E^j_{(b)}\right)\right|_\gamma,
\label{eq:AabRE}
\end{eqnarray}
where $\bar R$ describes the Riemann tensor for $\bar g_{ij}$. 
We note that the first equation follows the fact that $R_{UiU}^{~~~~j}$ for the metric \eqref{eq:orimet} does not depend on 
$a(U,V,Y^k)$ nor $b_i(U,V,Y^k)$. 
Since the PL metric in the Brinkmann coordinates is totally characterized by $A_{ab}$, 
we can obtain the metric form of Eq.~\eqref{eq:metBrink} by calculating 
the Riemann tensor components and the parallelly propagated zweibein in the metric \eqref{eq:orimet} 
without performing the explicit coordinate transformations.

\subsection{Construction of the Penrose coordinates based on the Hamilton-Jacobi function}
\label{sec:PLHJ}
The Hamilton-Jacobi equation for null geodesics is given by 
\begin{equation}
  g^{\mu\nu}\del_\mu S\del_\nu S=0. 
\end{equation}
The complete solution for the Hamilton-Jacobi equation contains arbitrary constants $\alpha_\mu$ as 
$S(x^\mu;\alpha^\mu)$. 
Once we set an initial hyper-surface satisfying $F(x^\mu_0)=0$ for an initial point $x^\mu=x^\mu_0$, 
the initial momentum can be calculated by $p_\mu^0=\del_\mu S|_{x^\mu=x_0^\mu}$. 
If the hypersurface intersects with each characteristic curve, we can construct a null geodesic congruence 
in the vicinity of a null geodesic $\gamma$. 
That is, a point in this congruence is specified by the initial point on the initial hyper-surface and the affine parameter $\tau$ as 
$x^\mu=x^\mu(\tau,x^\mu_0)$ with $F(x_0^\mu)=0$. 
Then we can introduce the new coordinates $(U,V,Y^k)$ as 
\begin{eqnarray}
  U&=&\tau,\\
  V&=&S(x^\mu_0), 
\end{eqnarray}
so that $x^\mu=x^\mu(U,x_0^\mu(V,Y^k))=x^\mu(U,V,Y^k)$. 
For this coordinate system, we find 
\begin{eqnarray}
  g_{UU}&=&g_{\mu\nu}\frac{dx^\mu}{d\tau}\frac{dx^\nu}{d\tau}=0, \\
  g_{UV}&=&g_{\mu\nu}\frac{\partial x^\mu}{\partial U}\frac{\partial x^\mu}{\partial V}=\partial_\mu S\frac{\partial x^\mu}{\partial V}=\frac{\partial S}{\partial V}=1, \\
  g_{Ui}&=&g_{\mu\nu}\frac{\partial x^\mu}{\partial U}\frac{\partial x^\mu}{\partial Y^i}=\frac{\partial S}{\partial Y^i}=\frac{\partial V}{\partial Y^i}=0. 
\end{eqnarray}
Therefore, the coordinates $(U, V, Y^k)$ provide an adapted coordinate system for the null geodesic $\gamma$. 

\section{Penrose limit for a circular photon orbit}
\label{sec:PLPS}

In this appendix, we follow this general procedure to construct a Penrose coordinate system adapted to a UCOP in a general static spherically symmetric spacetime. 
Before considering the Hamilton-Jacobi function, we need to change the angular coordinate describing the null geodesic motion. 
This is because the condition $\theta={\rm const.}$ cannot always be consistent with geodesic motion, but only for $\theta=\pi/2$. 
Therefore, the Hamilton-Jacobi function cannot be trivially extended to the region  $\theta\neq\pi/2$ in the conventional treatment. 
On the other hand, the condition $\phi={\rm const.}$ can always be consistent with the geodesic motion, and we adopt the angular coordinate $\theta$ instead of $\phi$ in this Appendix. 
The Lagrangian is given by 
\begin{equation}
  \mathcal L=\frac{1}{2}\left(-f\dot t^2+h\dot r^2+r^2\dot \theta^2\right). 
\end{equation}
The Hamilton-Jacobi function is written as 
\begin{equation}
  S(x^\mu)=-E t+L\theta+\mathcal \mathcal R (r). 
\end{equation}

In Ref.~\cite{Blau_review}, the coordinate construction is given for the Schwarzschild metric with 
the initial hyper-surface being $r={\rm const.}$ surface. 
However, this surface $r={\rm const.}$ is tangent to the characteristic curve $r=r_\rmc$, and does not match the case we suppose. 
Although the procedure shown in Ref.~\cite{Blau_review} gives the same 
PL metric in Brinkmann coordinates as ours, we avoid this subtlety by taking a different initial hyper-surface. 
Specifically, we implicitly assume that the initial hypersurface is given by $F(\theta,r)=0$. 

First, the coordinate $\phi$ is trivially related to the new coordinate $\tilde \phi$ by $\phi=\tilde \phi$. 
From our assumption for the initial hyper-surface, we consider the following coordinate transformations $(r,\theta)\rightarrow (U,\tilde r)$:
\begin{eqnarray}
  \dd\theta&=&\dot \theta \dd U+(\dot \theta-\beta)\dd \tilde r, \\
  \dd r&=& \dot r \dd U+\dot r \dd \tilde r, 
\end{eqnarray}
where the second terms have been fixed to guarantee the integrability and $\dd\theta=\dot \theta \dd U$ on the photon sphere. 
This coordinate transformation is singular for $\dot r=0$. 
Nevertheless, after the coordinate transformation to Brinkmann coordinates, this coordinate singularity can be resolved
\footnote{The coordinate singularity associated with the photon sphere, at which $\dot r=0$, cannot be avoided in the coordinate construction based on the Hamilton-Jacobi function due to the accumulation of the geodesic curves in the vicinity of the photon sphere. 
}. 
The coordinate transformation for $t$ is given by taking the derivative of the Hamilton-Jacobi function as 
\begin{eqnarray}
&&\dd S=\dd V=-E\dd t+L \dd \theta+R'(r)\dd r\\
&\Leftrightarrow&
\dd t=-\frac{1}{E}\dd V+\frac{\beta^2\Delta^2}{E^2}\dd U+\left[\frac{\Delta^2}{Efr}-\frac{L}{E}\left(\frac{L}{r^2}-\beta^2\right)\right]\dd \tilde r, 
\end{eqnarray}
where $r$ should be regarded as a function of $U$ and $\tilde r$. 
The metric form in terms of the adapted Penrose coordinates is given by 
\begin{eqnarray}
  \dd \bar s^2&=&2\dd U\dd V+\frac{\beta^2\Delta^2}{E^2}\dd \tilde r^2+r^2\sin^2\theta\dd \tilde \phi^2 \\
&&+{\rm (other~terms~involving~}\dd V{\rm )}, 
\end{eqnarray}
where $r$ and $\theta$ should be regarded as functions of $U$ and $\tilde r$. 
This metric form is preserved under the residual gauge transformation $U\rightarrow U+\Lambda(V,\tilde r,\tilde \phi)$ 
with $\Lambda$ being the arbitrary function of $V$, $\tilde r$ and $\tilde \phi$. 
The PL metric in Rosen coordinates associated with the above coordinate transformation is given by 
\begin{equation}
  \dd \bar s^2=2\dd U\dd V+\frac{\beta^2\Delta^2(r(U))}{E^2}\dd \tilde r^2+r^2(U)\sin^2\left(L\int r^{-2}(U)\dd U\right)\dd \tilde \phi^2,  
\end{equation}
where $r(U)=r(U,\tilde r=0)$. 
The zweibeins used in the coordinate transformation from Rosen to Brinkmann are given by 
\begin{equation}
  E_\mu^{(\tilde r)}=\frac{\beta\Delta(r(U))}{E}(\dd\tilde r)_\mu~~\&~~  
  E_\mu^{(\tilde \phi)}=r(U)\sin\left(L\int r^{-2}(U)\dd U\right)(\dd\tilde \phi)_\mu. 
\end{equation}
The metric components $A_{ab}$ can be easily calculated using Eq.~\eqref{eq:AabfromE}. 
In the limit approaching a UCOP, we obtain 
\begin{eqnarray}
  A_{11}&=&\left.\frac{\ddot \Delta}{\Delta}\right|_{r=r_\rmc}=-V_c''=\alpha^2, \\
  A_{22}&=&\left.\frac{(r\sin\theta)\ddot~}{(r\sin\theta)}\right|_{r=r_\rmc}=-\left.\dot\theta^2\right|_{r=r_\rmc}=-\beta^2. 
\end{eqnarray}
It would be worthwhile to explicitly write down the relation between the coordinates $(t,\theta)$ and $(u,v)$ 
on the UCOP. 
Including the gauge degree of freedom $\Lambda(v)$, we find 
\begin{eqnarray}
  t&\hat =&\frac{E}{f_\rmc}(u+\Lambda(v))-\frac{v}{E}, 
  \label{eq:tuv}\\
  \theta&\hat =&\beta(u+\Lambda(v)), 
  \label{eq:thuv}
\end{eqnarray}
where we have used the symbol $\hat =$ to indicate the equality only on the UCOP. 
These relations correspond to Eqs.~\eqref{eq:tuv0} and \eqref{eq:phuv}.


\end{document}